\documentclass[fontsize=10pt,headings=small]{scrartcl}
 \usepackage{a4wide}
\usepackage{amsmath}
\usepackage{amsfonts}
\usepackage{amssymb}
\usepackage{amsthm}
\usepackage{graphicx}
\usepackage[T1]{fontenc}
\usepackage[utf8]{inputenc}
\usepackage{booktabs}
\usepackage[english]{babel} 
\newtheorem{Definition}{Definition}
\newtheorem{Lemma}{Lemma}
\newtheorem{Proposition}{Proposition}
\newtheorem{Corollary}{Corollary}
\newtheorem{Remark}{Remark}
\theoremstyle{definition}
 
\usepackage{afterpage}
\usepackage{float}
\usepackage{multirow}
\usepackage{url}
\usepackage[colorinlistoftodos]{todonotes} 
\usepackage{color}

\author{
Roswitha Bammer\footnote{NuHAG, Faculty of Mathematics, Oskar-Morgensterin-Platz 1, 1090 Vienna}, Monika D\"orfler\footnotemark[1] and Pavol Harar\footnotemark[1] \footnote{Department of Telecommunications, Brno University of Technology, Brno, Czech Republic}}

\title{Gabor frames and deep scattering networks in audio processing}
\begin{document}
\maketitle
\section{Abstract}
\noindent
This paper introduces Gabor scattering, a~feature extractor based on Gabor frames and Mallat's scattering transform. By using a~simple signal model for audio signals, specific properties of Gabor scattering are studied. It is shown that, for each layer, specific invariances to certain signal characteristics occur. Furthermore, deformation stability of the coefficient vector generated by the  feature extractor is derived by using a~decoupling technique which exploits the contractivity of general scattering networks.  Deformations are introduced as changes in spectral shape and frequency modulation. The theoretical results are illustrated by numerical examples and experiments. Numerical evidence is given by evaluation on a~synthetic and a~``real'' dataset, that the invariances encoded by the Gabor scattering transform lead to higher performance in comparison with just using Gabor transform, especially when few training samples are available.





\section{Introduction}
\noindent
During the last two decades, enormous amounts of digitally encoded and stored audio have become available.
 For various purposes, the audio data, be it music or speech, need to be structured and understood. Recent  machine learning techniques, known as (deep) convolutional neural networks (CNN), have led to state of the art results for several tasks such as classification, segmentation or voice detection, cf. \cite{NIPS2012_4824,  grill2015_ismir}.
CNNs were originally proposed for images \cite{NIPS2012_4824}, which may be directly fed into a~network. Audio signals, on the other hand, commonly undergo some preprocessing to extract features that are then used as input to a~trainable machine. Very often, these features consist of one or several two-dimensional arrays, such that the image processing situation is mimicked in a~certain sense. However, the question about the impact of this very first processing step is important and it is not entirely clear whether a~short-time Fourier transform (STFT), here based on \textit{Gabor frames}, the most common representation system used in the analysis of audio signals, leads to optimal feature extraction. 
The convolutional layers of the CNNs can themselves be seen as feature extractors,
often followed by a~classification stage, either in the form of one or several dense network
layers or classification tools such as support vector machine (SVM). St{\'e}phane Mallat
gave a~ first mathematical analysis of CNN as feature extractor, thereby introducing the
so called \textit{scattering transform}, based on wavelet transforms and modulus
nonlinearity in each layer~\cite{mallat}. The basic structure thus parallels the one of
CNNs, as these networks are equally composed of multiple layers of local
convolutions,  followed by a~nonlinearity and, optionally, a~pooling operator, cp.,
Section \ref{DCNN}. 

\newpage
In the present contribution, we consider an approach inspired by Mallat's scattering
transform, but based on Gabor frames, respectively, Gabor transform (GT). The resulting feature extractor is called 
Gabor scattering (GS).  Our approach is a~special case of the extension of Mallat's scattering transform proposed by Wiatowski and
B\"olcskei~\cite{DBLP:journals/corr/WiatowskiB15a, deepwindow}, which introduces the
possibility to use different semi-discrete frames, Lipschitz-continuous nonlinearities
and pooling operators in each layer.
\mbox{In~\cite{mallat, Anden14, anloma15}}, invariance and deformation stability properties of
the scattering transform with respect to operators defined via some group action were
studied.  In the more general setting of~\cite{DBLP:journals/corr/WiatowskiB15a, deepwindow}, vertical translation invariance, depending on the network depth, and
deformation stability for band-limited functions have been proved.
In this contribution, we study the same properties of the GS and a~particular class of signals, which model simple musical tones (Section~\ref{model}). 

Due to this concrete setting, we obtain quantitative invariance statements and deformation stability to specific, musically meaningful, signal deformations.   
Invariances are studied considering the first two layers, where the feature extractor extracts certain signal features of the signal model (i.e.,~frequency and envelope information), cp., Section~\ref{scatmodel}.
By using a~low-pass filter and pooling in each layer, the temporal fine structure of the signal is averaged out. This results in invariance with respect to the envelope in the first and frequency invariance in the second layer output.
To compute deformation bounds for the GS feature extractor, we assume more specific restrictions than band-limitation and use the decoupling technique, first presented in~\cite{DBLP:journals/corr/WiatowskiB15a,7541482}. Deformation stability is proven by only computing the robustness of the signal class w.r.t spectral shape and frequency modulation, see Section~\ref{deform}. The robustness results together with contractivity of the feature extractor, which is given by the networks architecture, yields deformation stability.

To empirically demonstrate the benefits of GS time-frequency representation for classification, we have conducted a~set of experiments. In a~supervised learning setting, where the main aim is the multiclass classification of generated sounds, we have utilized a~CNN as a~classifier.
In these numerical experiments, we compare the GS to a~STFT-based representation. We demonstrate the benefits of GS in a~quasi-ideal setting on a~self implemented synthetic dataset, and we also investigate if it benefits the performance on a~real dataset, namely, GoodSounds \cite{GoodSounds}.  Moreover we focus on comparing these two time-frequency representations in terms of performance on limited sizes of training data, see Section \ref{section:expresults}.

\subsection{Convolutional Neural Networks (CNNs) and Invariance}
\noindent
\label{DCNN}
{CNNs are a~specific class of neural network architectures which have shown
extremely convincing results on various machine learning tasks in the past decade. Most
of the problems addressed using CNNs are based on, often, big amounts of annotated
data, in which case one speaks about supervised learning. When learning from data, the
intrinsic task of the learning architecture is to gradually extract useful information
and suppress redundancies, which always abound in natural data. More formally, the~learning problem of interest may be invariant to various changes of the original data
and the machine or network must learn these invariances in order to avoid overfitting.
As, given a~sufficiently rich architecture, a~deep neural network can practically fit arbitrary data, cp.~\cite{45820, kawaguchi2017generalization}, good generalization properties depend on the systematic incorporation of the intrinsic invariances of the data. 
 Generalization properties hence suffer if the architecture is too rich given the amount of
 available data. This problem is often addressed by using data augmentation.  Here, \textit{ we raise the
 hypothesis that using prior representations which encode some potentially useful invariances  will increase the generalization quality, in particular when using a~restricted size of data
 set.}  The evaluation of  the performance on validation data in comparison to the results on
 test data strengthens our hypothesis for the experimental problem presented in Section~\ref{section:expresults}.}

{To understand the mathematical construction used within this paper, we briefly introduce the principal idea and structure of a~CNN. We shall see that the scattering transforms, in general, and~the GS, in particular, follow a~similar concept of concatenating various processing steps, which ultimately leads to rather flexible grades of invariances in dependence on the chosen parameters. 
Usually, a~CNN consists of several layers, namely, an input, several hidden
(as we consider the case of deep 
CNN the number of hidden layers 
is supposed to be $\geq$2) and one output layer. A~hidden layer consists
of the following steps: first the convolution of the data with a~small
weighting matrix, often~referred to as a~kernel\footnote{We~point out that the term kernel as used in this work always means convolutional kernels in the sense of filterbanks. Both the fixed kernels used in the scattering transform and the kernels used in the CNNs, whose size is fixed but whose elements are learned, should be interpreted as convolutional kernels in a filterbank. This~should not be confused with the kernels used in classical machine learning methods based on reproducing kernel Hilbert spaces, e.g., the famous support vector machine, c.f.~\cite{HoSc08}}, which~can be interpreted as 
localization of certain properties of the input data.} 
The main advantage of this setup is that only the size and number of these (convolutional) kernels are fixed, but their coefficients are learned during training.  So they reflect the structure of the training data in the best way w.r.t the task being solved.
The next building block of the hidden layer is the application of a~nonlinearity function, also called activation function, which signals if information of this neuron is relevant to be transmitted. Finally, to reduce redundancy and increase invariance, pooling is applied. Due to these building blocks, invariances to specific deformations and variations in the dataset are generated in dependence on the specific filters used, whether they are learned, as in the classical CNN case, or designed, as in the case of scattering transforms \cite{ma16}. In this work, we will derive concrete qualitative statements about invariances for a~class of music signals and will show by numerical experiments that these invariances indeed lead to a~better generalization of the CNNs used to classify data. 

Note that in  a~neural network, in particular in  CNNs, the output, e.g., classification labels,  is~obtained after several concatenated hidden layers. In the case of scattering network, the outputs of each layer are stacked together into a~feature vector, and further processing is necessary to obtain the desired result. Usually, after some kind of dimensionality reduction, cf. ~\cite{Wietal16}, this vector can be fed into a~SVM or a~dense NN, which performs the classification task.

\subsection{Invarince Induced by Gabor Scattering}
\label{inv_gs}

\noindent

\begin{figure}[H]
\centering
 \includegraphics[width=3.5in]{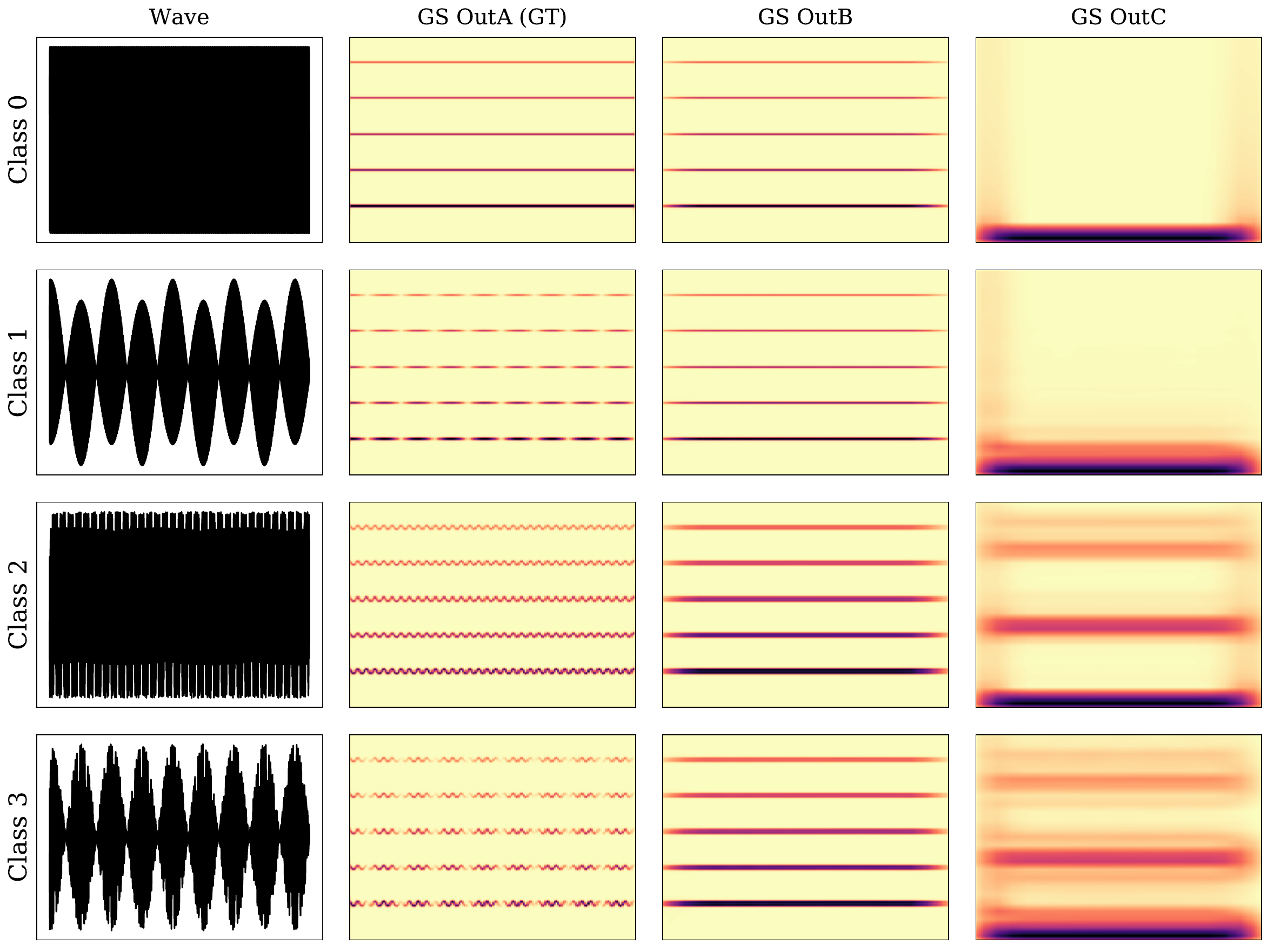}
\caption{Wave Outputs~A, i.e., GT; B; and C of Gabor scattering (GS) for all four classes of generated~sound.}
\label{fig:visualization_synth}
\end{figure}

In this section, we give a~motivation 
for the theory and  underlying aim of this paper. 
In Figure \ref{fig:visualization_synth}, we
see sound examples from  different
classes, where Class~0 is a~pure tone
with 5~harmonics, Class~1 is an amplitude
modulated version thereof, Class~2 is the frequency modulated version, and Class~3 contains the amplitude
and frequency modulated signal. So, we have 
classes with different amplitudes, as~is clearly visible in 
the waveforms shown in the left-most plots. In this paper, we 
introduce GS, as~a~new feature extractor that introduces certain 
invariances. GS has several layers, 
denoted by  OutA, OutB, and OutC, and each layer is invariant 
with respect to some features. The first layer, here~OutA, is~the spectrogram of the waveform. So, we see the 
time-frequency content of the four classes. OutB~can be seen 
to be  invariant with respect to  amplitude changes, whereas the last 
layer, OutC, is invariant with respect to to  frequency content while encoding the amplitude information.
With GS it is therefore possible to separate different qualities of  information contained in a~spectrogram.

We introduce GS mathematically in Section \ref{basic} and elaborate on the resulting invariances in different layers in Section \ref{scatmodel}.  Numerical experiments, showing the benefit of GS, are discussed  in Section \ref{section:expresults}.

\section{Materials and Methods}
\subsection{Gabor Scattering} \label{basic}
\noindent
As Wiatowski and B\"olcskei used general semi-discrete frames to obtain a~wider class of window functions for the scattering transform (cp. \cite{DBLP:journals/corr/WiatowskiB15a, deepwindow}), it seems natural to consider specific frames used for audio data analysis. Therefore, we use Gabor frames for the scattering transform and study corresponding properties. We next introduce the basics of Gabor frames, refer to  \cite{opac-b1098011} for more details.
A sequence $(g_k)_{k=1}^{\infty}$ of elements in a~Hilbert space $\mathcal{H}$ is called frame if there exist positive frame bounds $A, B>0$ such that for all $f \in \mathcal{H}$
\begin{equation}
A \| f\|^2 \leq \sum_{k=1}^{\infty} |\langle f, g_k\rangle |^2\leq B \| f\| ^2 .
\label{eq:Gabor}
\end{equation}
 
 If $A=B,$ then we call $(g_k)_{k=1}^{\infty}$ a~tight frame. 
\begin{Remark}
In our context, the Hilbert space $\mathcal{H}$ is either $L^2(\mathbb{R})$ or $\ell^2(\mathbb{Z}).$
\end{Remark}
\noindent
To define Gabor frames we need to introduce two operators, i.e., the translation and modulation operator. 
\begin{itemize}
\item The translation (time shift) operator: 
\begin{itemize}
    \item for a~function $f \in L^2(\mathbb{R})$ and $x \in \mathbb{R}$ is defined as $T_x f(t):=f(t-x)$ for all $t \in\mathbb{R}.$
    \item for a~function $f \in \ell^2(\mathbb{Z})$ and $ k \in \mathbb{Z}$ is defined as $T_{k} f(j):=(f(j-k))_{j \in \mathbb{Z}}.$ 
\end{itemize}

\item The modulation (frequency shift) operator: 
\begin{itemize}
    \item for a~function $f \in L^2(\mathbb{R})$ and $\omega \in \mathbb{R}$ is defined as $M_\omega f(t):=e^{2\pi i  \omega t}f(t) $ for all $t \in \mathbb{R}.$

    \item for a~function $f \in \ell^2(\mathbb{Z})$ and $ \omega \in[-\frac{1}{2},\frac{1}2{}]$ is defined as $M_\omega f(j):=(e^{2\pi i \omega j}f(j))_{j \in \mathbb{Z}}.$
\end{itemize}

\end{itemize}

We use these operators to express the STFT of a~function $f\in \mathcal{H}$ with respect to a~given window function $g\in \mathcal{H}$ as
$V_gf(x,\omega)=\langle f,M_\omega T_x g \rangle.$
To reduce redundancy, we  sample $V_gf$ on a~separable lattice $\Lambda= \alpha\mathbb{Z}\times \mathcal{I}$, 
where $\mathcal{I}= \beta\mathbb{Z}$ in case of $\mathcal{H} = L^2(\mathbb{R})$, and $\mathcal{I}= \{ 0, ..., \frac{(M-1)}{M}\}$ with $\beta = \frac{1}{M}$ in case $\mathcal{H} = \ell^2(\mathbb{Z}).$
The sampling is done in time by $\alpha>0$ and in frequency by $\beta>0.$ 
The~resulting samples correspond to the coefficients of $f$ with respect to a~``Gabor system''.
\begin{Definition}
\emph{\textbf{(Gabor System)}
Given} a~window function $0\neq g \in \mathcal{H}$ and lattice parameters $\alpha, \beta>0$, the set of time-frequency shifted versions of $g$
$$\mathit{G}(g,\alpha,\beta)=\{ M_{\beta j}T_{\alpha k}g : (\alpha k, \beta j)\in \Lambda\}$$ 
is called a~Gabor system. 
\end{Definition}

This Gabor system is called Gabor frame if it is a~frame, see Equation (\ref{eq:Gabor}).
We proceed to introduce a~scattering transform based on Gabor frames. We base our considerations on  \cite{DBLP:journals/corr/WiatowskiB15a} by using a~triplet sequence $\Omega = \big{(} (\Psi_\ell,\sigma_\ell,S_\ell)\big{)}_{\ell \in \mathbb{N}},$ where $\ell$ is associated to the $\ell$-th layer of the network.
Note that in this contribution, we will deal with Hilbert spaces $L^2(\mathbb{R})$ or $\ell^2 (\mathbb{Z})$; more precisely in the input layer, i.e., the $0-$th layer, we have $\mathcal{H}_0 = L^2(\mathbb{R})$ and, due to the discretization inherent in the GT, $\mathcal{H_\ell} =  \ell^2 (\mathbb{Z}) \,\,\, \forall \ell>0.$

We recall the elements of the triplet: 
\begin{itemize}
\item $\Psi_\ell := \{ g_{\lambda_\ell}\}_{\lambda_\ell \in \Lambda_\ell}$ with $g_{\lambda_\ell} = M_{\beta_\ell j} T_{\alpha_\ell k} g_\ell,$ $\lambda_\ell = (\alpha_\ell k,\beta_\ell j),$  is a~Gabor frame indexed by a~lattice~$\Lambda_\ell.$ 
\item  A~nonlinearity function (e.g., rectified linear units, modulus function, see  \cite{DBLP:journals/corr/WiatowskiB15a}) $\sigma_\ell: \mathbb{C}\to \mathbb{C},$ is~applied pointwise and  is chosen to be Lipschitz-continuous, i.e., $\| \sigma_\ell f-\sigma_\ell h\|_2 \leq L_\ell \| f-h\|_2$ for all $f,h \in \mathcal{H}.$ In this paper we only use the modulus function with Lipschitz constant $L_\ell = 1$ for all $\ell \in \mathbb{N}.$
\item Pooling depends on a~pooling factor $S_\ell >0,$ which leads to dimensionality reduction.  Mostly used are max- or average-pooling, some more examples can be found in  \cite{DBLP:journals/corr/WiatowskiB15a}.
In our context, pooling is covered by choosing specific lattices $\Lambda_\ell$ in each layer.
\end{itemize}
\noindent
To explain the interpretation of GS as CNN, we write
 $\mathcal{I}(g)(t) = g(-t)$ and have
\begin{align}
|\langle f, M_{\beta j} T_{\alpha k}g\rangle | =\left|f \ast \left(\mathcal{I}\big{(}M_{\beta j}(g)\big{)}\right)\right| (\alpha k).
\label{invol}
\end{align}
Thus, the Gabor coefficients can be interpreted as the samples of a~convolution.

We start by defining ``paths'' on index sets $q := (q_1,...,q_\ell ) = ( \beta_1 j_1 , ...,  \beta_\ell j_\ell ) \in \beta_1\mathbb{Z} \times ... \times \beta_\ell \mathbb{Z} =: \mathcal{B}^\ell, \ell \in \mathbb{N}.$

\begin{Definition}
\emph{(Gabor Scattering)}
Let  $\Omega = \big{(} (\Psi_\ell,\sigma_\ell,\Lambda_\ell)\big{)}_{\ell \in \mathbb{N}}$ be a~given triplet sequence.
Then, the components of  the $\ell $-th layer of the GS transform are  defined to be  the output of the operator 
$U_\ell[q_\ell]: \mathcal{H}_{\ell-1} \to \mathcal{H}_{\ell}$, $q_\ell\in  \beta_\ell \mathbb{Z}$:
\begin{align}
f_\ell^{(q_1,...,q_\ell)}(k)=U_\ell[\beta_\ell j_\ell] f_{\ell-1}^{(q_1,...,q_{\ell-1})} (k):= \sigma_\ell \left(\langle f_{\ell-1}^{(q_1,...,q_{\ell-1})}, M_{\beta_\ell j_\ell}T_{\alpha_\ell k}g_\ell\rangle_{\mathcal{H}_{\ell-1}}\right)\,\,\,\, j_\ell,k \in \mathbb{Z},
\label{eq:nlayer}
\end{align}
where $f_{\ell-1}$ is some output-vector of the previous layer and $f_\ell \in \mathcal{H}_\ell \,\, \, \forall \ell \in \mathbb{N}.$
The GS operator is defined as
$$ U[q]f = U[(q_1,...,q_\ell)]f := U_\ell[q_\ell]\cdot \cdot \cdot U_1[q_1]f.$$
\end{Definition}
\noindent
Similar to  \cite{DBLP:journals/corr/WiatowskiB15a}, for each layer, we use one atom of the Gabor frame in the subsequent layer as output-generating atom, i.e., $\phi_{\ell-1} := g_{\ell}.$ Note that convolution with this element corresponds to low-pass filtering \footnote{In general, one could take $\phi_{\ell-1} := g_{\lambda_\ell^*}, \lambda_\ell^* \in \Lambda_\ell.$ As this element is the $\ell$-th convolution, it is an element of the $\ell$-th frame, but because it belongs to the $(\ell-1)$-th layer, its index is $(\ell-1)$.}. 
We next introduce a~countable set $\mathcal{Q} : =  \bigcup_{\ell=0}^{\infty} \mathcal{B}^\ell,$  which is the union of all possible paths of the net and the space $(\ell^2(\mathbb{Z}))^\mathcal{Q}$ of sets of $\mathcal{Q}$ elements from  $ \ell^2(\mathbb{Z}).$
Now we define the feature extractor $\Phi_\Omega (f)$ of a~signal $f \in L^2(\mathbb{R})$ as in (\cite{DBLP:journals/corr/WiatowskiB15a}, Definition \ref{Def3}) based on chosen (not learned) Gabor~windows.
\begin{Definition}\label{Def3}
\emph{(Feature Extractor)}
Let  $\Omega = \big{(} (\Psi_\ell,\sigma_\ell,\Lambda_\ell)\big{)}_{\ell \in \mathbb{N}}$ be a~triplet sequence and $\phi_{\ell}$ the output-generating atom for layer $\ell$.
Then the feature extractor $\Phi_\Omega: L^2(\mathbb{R}) \to ( \ell^2(\mathbb{Z}))^\mathcal{Q}$  is defined as
\begin{align}
\Phi_\Omega (f) : = \bigcup_{\ell=0}^{\infty} \left\{ ( U[q]f) \ast \phi_\ell \right\} _{q\in \mathcal{B}^\ell}.
\label{eq:output}
\end{align}
\end{Definition}
\noindent
{In the following section we are going to introduce the signal model which we consider in this paper}.

\subsection{Musical Signal Model} \label{model}
\noindent
Tones are one of the smallest units and simple models of an audio signal, consisting of one fundamental frequency $\xi_0$, corresponding harmonics $n\xi_0$, and a~shaping envelope $A_n$ for each harmonic, providing specific timbre. 
Further, as our ears are limited to frequencies below 20 kHz, we develop our model over finitely many harmonics, i.e., $\{1,...,N\}\subset\mathbb{N}$.

The general model has the following form,
\begin{align}
f(t) = \sum_{n=1}^N A_n(t) e^{2 \pi i \eta_n(t)}, \label{eq:model}
\end{align}
where $A_n(t) \geq 0 \,\,\, \forall n \in \{1,...,N\}$ and $\forall t.$
For one single tone we choose $\eta_n(t) = n\xi_0 t.$ 
Moreover, we create a~space of tones $\mathcal{T} = \big{\{} \sum_{n=1}^N A_n(t) e^{2 \pi i n \xi_0 t} | A_n \in\mathcal{C}^\infty_c (\mathbb{R})\big{\}}$ and assume $\| A_n \|_\infty \leq \frac{1}{n}.$ 
 

\section{Theoretical Results}
\vspace{-6pt} 
\subsection{Gabor Scattering of Music Signals}
\vspace{-6pt}
\subsubsection{Invariance}
\label{scatmodel}
\noindent
In  \cite{Anden14}, it was already stated that due to the structure of the scattering transform the energy of the signal is pushed towards low frequencies,
where it is then captured by a~low-pass filter as output-generating atom. The current section explains how GS separates relevant structures of signals modeled by the signal space $\mathcal{T} $. 
Due to the smoothing action of the output-generating atom, each~layer expresses certain invariances, which will be illustrated by numerical examples in Section~\ref{example}. 
In~Proposition \ref{prop}, inspired by \cite{Anden14}, we add some assumptions on the analysis window in the first layer $g_1: |\hat{g}_1(\omega)| \leq C_{\hat{g}_1} (1+|\omega|^s)^{-1}$ for some $s>1$ and $\|t g_1(t)\|_1= C_{g_1} < \infty.$

\begin{Proposition}[Layer $1$]
\label{prop}
Let $f \in \mathcal{T}$ with $\|A'_n \|_\infty \leq C_n< \infty$   $ \forall n\in \{1,...,N\}.$ 
For fixed j, for which  $n_0 =\underset{n\in \{1,...,N\}}{ \mbox{argmin}} |\beta_1 j - \xi_0 n|$ such that  $|\beta j -\xi_0 n_0|\leq \frac{\xi_0}{2},$ can be found, we obtain 
\begin{align}
&U [\beta_1 j](f)(k)=|\langle f, M_{\beta_1 j}T_{\alpha_1 k}g_1\rangle| = A_{n_0}(\alpha_1 k)| \hat{g}_1(\beta_1 j- n_0\xi_0)| + E_1( k) \label{eq:prop}\\
&E_1( k) \leq  C_{g_1}  \sum_{n=1}^N \| A'_n \cdot T_k\chi[-\alpha_1; \alpha_1]\|_\infty + C_{\hat{g}_1} \sum_{n = 2-n_0}^{N-n_0}  \frac{1}{n_0+n-1}\left(1+ \Big{|}\xi_0\Big{|}^s \Big|n-\frac{1}{2}\Big|^s \right)^{-1},
\label{eq:prop1}
\end{align}
where $\chi$ is the indicator function.
\end{Proposition}
\noindent
\begin{Remark}
Equation \eqref{eq:prop} shows that for slowly varying amplitude functions $A_n$, the first layer mainly captures the contributions near the frequencies of the tone's harmonics. Obviously, for time sections during which the envelopes $A_n$ undergo faster changes, such as during a~tone's onset, energy will also be found outside a~small interval around the harmonics' frequencies and thus the error estimate Equation \eqref{eq:prop1} becomes less stringent. The~second term of the error in Equation \eqref{eq:prop1} depends only on the window $g_1$ and its behavior is governed by the frequency decay of $g_1$. Note that the error bound increases for lower frequencies, as the separation of the fundamental frequency and corresponding harmonics by the analysis window deteriorates. 
\end{Remark}

\begin{proof} 
\textit{Step 1}: Using the signal model for tones as input, interchanging the finite sum with the integral and performing a~substitution $u = t - \alpha_1 k,$ we obtain
\begin{align}
\langle f, M_{\beta_1 j}T_{\alpha_1 k}g_1\rangle &= \langle \sum_{n =1}^N M_{n \xi_0}A_n , M_{\beta_1 j}T_{\alpha_1 k}g_1\rangle \notag\\
&= \sum_{n=1}^N \langle A_n, M_{\beta_1 j- n \xi_0} T_{\alpha_1 k} g_1 \rangle \notag \\
& = \sum_{n =1}^N \int_\mathbb{R} A_n(u+\alpha_1 k) g_1(u) e^{-2 \pi i (\beta_1 j-n \xi_0 )(u+\alpha_1 k)} du.\notag
\end{align}

After performing a~Taylor series expansion locally around $\alpha_1 k:\,$\\ $A_n(u+\alpha_1 k) = A_n(\alpha_1 k) + u R_n(\alpha_1 k,u),$ where the remainder can be estimated by $|R_n(\alpha_1 k,u)| \leq \| A'_n \cdot T_k\chi[-\alpha_1; \alpha_1]\|_\infty, $
we have
\begin{align}
&\langle f, M_{\beta_1 j}T_{\alpha_1 k}g_1\rangle =\sum_{n=1}^N \bigg{[} e^{-2 \pi i(\beta_1 j-n \xi_0 )\alpha_1 k}A_n(\alpha_1 k) \int_\mathbb{R} g_1(u) e^{-2 \pi i(\beta_1 j-n \xi_0 )u} du\notag
\\
&+\int_\mathbb{R} u R_n(\alpha_1 k,u)g_1(u) e^{-2 \pi i(\beta_1 j-n \xi_0 )(u+\alpha_1 k)} du\bigg{]}. \notag 
\end{align}

Therefore, we choose $n_0 =\underset{n}{ \mbox{argmin}} |\beta_1 j- \xi_0 n|,$ set 
\begin{align}
&\mathcal{E}_n(k) =\!\! \int_\mathbb{R} \! \! u R_n(\alpha_1 k,u)g_1(u) e^{-2 \pi i(\beta_1 j-n \xi_0 )(u+\alpha_1 k)} du \label{eq:bound1}\\
&\widetilde{E}( k) = \sum_{\substack{n=1\\n \ne n_o}}^N e^{-2 \pi i(\beta_1 j-n \xi_0 )\alpha_1 k}A_n(\alpha_1 k) \hat{g}_1(\beta_1 j- n\xi_0)\label{eq:bound2}
\end{align}
and split the sum to obtain
\begin{align}
&\langle f, M_{\beta_1 j}T_{\alpha_1 k}g_1\rangle = A_{n_0}(\alpha_1 k) e^{-2 \pi i( \beta_1 j-n_0 \xi_0 )\alpha_1 k}\hat{g}_1(\beta_1 j- n_0\xi_0) +  \widetilde{E}( k) + \sum_{n=1}^N  \mathcal{E}_n(k). \notag
\end{align} 

\textit{Step~2}: We bound the error terms, starting with Equation \eqref{eq:bound1}:

$$\left|\sum_{n=1}^N \mathcal{E}_n( k)\right| = \left| \sum_{n=1}^N \int_\mathbb{R} u R_n(\alpha_1 k,u)g_1(u) e^{-2 \pi i( \beta_1 j-n_0 \xi_0 )(u+\alpha_1 k)} du\right|.$$

Using triangle inequality and the estimate for the Taylor remainder, we obtain, together with the assumption on the analysis window,
\begin{align}
\left|\sum_{n=1}^N \mathcal{E}_n( k)\right|&\leq \sum_{n=1}^N \| A'_n \cdot T_k\chi[-\alpha_1; \alpha_1]\|_\infty \int_\mathbb{R}| u g_1(u) |du  \notag \\
&\leq C_{g_1} \sum_{n=1}^N \| A'_n \cdot T_k\chi[-\alpha_1; \alpha_1]\|_\infty. \notag
\end{align}

For the second bound, i.e., the bound of Equation \eqref{eq:bound2}, we use the decay condition on $\hat{g}_1,$ thus
\begin{align}
| \widetilde{E}(k)| \leq C_{\hat{g}_1} \sum_{\substack{n=1\\n \ne n_o}}^N \left|A_n(\alpha_1 k)\right| \big{(}1+|\beta_1 j -\xi_0 n|^s\big{)}^{-1}.\notag
\end{align}

Next we split the sum into $n>n_0$ and $n< n_0.$ We estimate the error term for $n>n_0\!:$

\begin{align}
\sum_{n=n_0+1}^N |A_n(\alpha_1 k)| \big{(}1+|\beta_1 j -\xi_0 n|^s\big{)}^{-1} = \sum_{n=1}^{N-n_0} |A_{n_0+n}(\alpha_1 k)| \big{(}1+|\beta_1 j -\xi_0 n_0-\xi_0n|^s\big{)}^{-1}. 
\end{align}

As $n_0 =\underset{n}{ \mbox{argmin}} |\beta_1 j - \xi_0 n|,$ we have  $|\beta_1 j- \xi_0n_0 | \leq \frac{\xi_0}{2}$ and, also, using $\| A_n \|_\infty \leq \frac{1}{n},$ we obtain
\begin{align}
&  \sum_{n=1}^{N-n_0} |A_{n_0+n}(\alpha_1 k)| \left(1+\Big|\frac{\xi_0}{2}-\xi_0 n \Big|^s\right)^{-1}  \leq \sum_{n=1}^{N-n_0} \frac{1}{n_0+n}\left( 1+\Big{|}\xi_0\Big{|}^s \Big|n-\frac{1}{2} \Big|^s\right)^{-1}.
\label{eq:nbig}
\end{align}

Further we estimate the error for $n<n_0\!:$
 \begin{align}
\sum_{n=1}^{n_0-1} |A_n(\alpha_1 k)| (1+|\beta_1 j -\xi_0 n|^s)^{-1} \leq \sum_{n=1}^{n_0-1} |A_n(\alpha_1 k)| (1+|\beta_1 j -\xi_0 n_0+\xi_0 n_0-\xi_0 n|^s)^{-1},\notag
\end{align}
where we added and subtracted the term $\xi_0 n_0.$
Due to the reverse triangle inequality and $|\beta_1 j- \xi_0n_0 | \leq \frac{\xi_0}{2}$, we obtain
\begin{align}
    \Big{|}\beta_1 j - \xi_0n_0 - \xi_0 (n-n_0)\Big{|}\geq \Big{|}\xi_0 (n_0-n) - \frac{\xi_0}{2}\Big{|}.\notag
\end{align}

For convenience, we call $m = n- n_0$ and perform a little trick by adding and subtracting $\frac{1}{2}$, so~$\Big{|}\xi_0 (n_0-n) - \frac{\xi_0}{2}\Big{|} =|\xi_0|\Big{|}-( m+1)   + \frac{1}{2}\Big{|}.$ The reason for this steps will become more clear when putting the two sums back together.
Now, we have
\begin{align}
\sum_{n=1}^{n_0-1} |A_n(\alpha_1 k)| (1+|\beta_1 j -\xi_0 n|^s)^{-1} \leq \sum_{m=1-n_0}^{-1} |A_{n_0+m}(\alpha_1 k)| \Big{(}1+\Big{|}\xi_0\Big{|}^s\Big{|}(m+1)-\frac{1}{2}\Big{|}^s\Big{)}^{-1}.\notag
\end{align}

Shifting the sum, i.e., taking $n = m+1,$ and using $\| A_n \|_\infty \leq \frac{1}{n},$ we get 
\begin{align}
\sum_{m=1-n_0}^{-1} |A_{n_0+m}(\alpha_1 k)| \Big{(}1+\Big{|}\xi_0\Big{|}^s\Big{|}(m+1)-\frac{1}{2}\Big{|}^s\Big{)}^{-1}\leq
     \sum_{n=2-n_0}^{0} \frac{1}{n_0+n-1} \Big{(}1+\Big{|}\xi_0\Big{|}^s\Big{|}n-\frac{1}{2}\Big{|}^s\Big{)}^{-1}.
     \label{eq:nlow}
\end{align}

Combining the two sums Equations \eqref{eq:nbig} and \eqref{eq:nlow} and observing that $\frac{1}{n_0+n}<\frac{1}{n_0+n-1}$, we obtain
\begin{align}
| \widetilde{E}(k)| \leq C_{\hat{g}_1} \sum_{n = 2-n_0}^{N-n_0}\frac{1}{n_0+n-1} \left(1+ \Big{|}\xi_0\Big{|}^s \Big|n-\frac{1}{2}\Big|^s \right)^{-1}. \label{eq:freqerror}
\end{align}

Summing up the error terms, we obtain Equation \eqref{eq:prop1}.
\begin{flushright}
\qedhere
\end{flushright}
\end{proof}

\noindent
To obtain the GS coefficients, we need to apply the output-generating atom as in Equation \eqref{eq:output}.
\begin{Corollary}[Output of Layer $1$]
\label{Cor1}
Let $\phi_1 \in \Psi_2$ be the output-generating atom, then the output of the first layer is 
\begin{align}
\big{(}U_1[\beta_1 j]f \ast \phi_1\big{)} (k)= |\hat{g}_1(\beta_1 j- n_0\xi_0)|(A_{n_0} \ast \phi_1)(k)+ \epsilon_1(k), \notag
\end{align}
where
 $$\epsilon_1(k) \leq \|E_1\|_\infty^2 \|\phi_1\|_1^2.$$
 
 Here $E_1$ is the error term of Proposition \ref{prop}.
\end{Corollary}

\begin{Remark} \label{rem1}
Note that we focus here on an unmodulated Gabor frame element $\phi_1$, and the convolution may be interpreted as a~low-pass filter.  Therefore, in dependence  on the pooling factor $\alpha_1$,  the temporal fine-structure of $A_{n_0}$ corresponding to higher frequency content is averaged out. 
\end{Remark}

\begin{proof}
For this proof, we use the result of Proposition \ref{prop}. We show that the calculations for the first layer are similar to~those of the second layer:
\begin{align}
&\bigg{|}\sum_k \big{(}| \langle f, M_{\beta_1 j} T_{\alpha_1 k}g_1\rangle|  -  |\hat{g}_1(\beta_1 j -\xi_0 n_0)|  A_{n_0}(k)\big{)}\cdot\phi_1(l-k)\bigg{|}^2 \\
&= \big{|}\sum_k E_1( k) \phi_1(l-k) \big{|}^2 \leq \|E_1\|_\infty^2 \|\phi_1\|_1^2 \notag
\end{align}
where $E_1( k) \leq  C_{g_1}  \sum_{n=1}^N \| A'_n \cdot T_k\chi[-\alpha_1; \alpha_1]\|_\infty + C_{\hat{g}_1} \sum_{n = 2-n_0}^{N-n_0}   \frac{1}{n_0+n-1}\left(1+ |\xi_0|^s |n-\frac{1}{2}|^s \right)^{-1}.$
\begin{flushright}
\qedhere
\end{flushright}
\end{proof}
\noindent
We introduce two more operators, first the sampling operator $S_\alpha\big{(}f(x)\big{)} = f(\alpha x)$\\$ \forall x\in \mathbb{R}$ and second the periodization operator $P_{\frac{1}{\alpha}}\big{(}\hat{f}(\omega)\big{)}=\sum_{k\in \mathbb{Z}}\hat{f}(\omega-\frac{k}{\alpha}) \,\,\, \forall \omega\in \mathbb{R}.$ These operators have the following relation $ \mathcal{F}\big{(}S_\alpha(f)\big{)}(\omega) =P_{\frac{1}{\alpha}}(\hat{f}(\omega)) .$
In order to see how the second layer captures relevant signal structures, depending on the first layer, we propose the following Proposition \ref{cor}. Recall that $g_\ell \in \mathcal{H}_\ell \,\, \forall \ell \in \mathbb{N}.$ 
\begin{Proposition}[Layer $2$]
\label{cor}
Let $f \in \mathcal{T}, $ $\sum_{k \neq 0} | \hat{A}_{n_0}(.-\frac{k}{\alpha_1})| \leq \varepsilon_{\alpha_1}$ and $ |\hat{g}_2(h)|\leq C_{\hat{g}_2} (1+|h|^s)^{-1}.$  Then the elements of the second layer can be expressed as
\begin{align}
U_2[\beta_2 h] U_1[\beta_1j]f(m) =  \big{|}\hat{g}_1(\beta_1 j -\xi_0 n_0)\big{|} \big{|}\big{\langle}M_{-\beta_2 h} A_{n_0}, T_{\alpha_2 m} g_2\big{\rangle}\big{|}+ E_2( m), 
\label{eq:PropLayer2}
\end{align}
where $$E_2( m)\leq \varepsilon_{\alpha_1} C_{\hat{g}_2}   |\hat{g}_1(\beta_1 j -\xi_0 n_0)|\sum_r \big{(} 1+|\beta_2h-r|^s\big{)}^{-1}+ \|E_1\|_\infty \cdot  \|g\|_1.$$
\end{Proposition}
\noindent

\begin{Remark}
Note that, as the envelopes $A_{n}$ are expected to change slowly except around
transients, their~Fourier transforms concentrate their energy in the low frequency
range.
Moreover, the modulation term $M_{-\beta_2 h}$ pushes the frequencies of $A_{n_0}$ down by $-\beta_2 h$, and therefore they can be captured by the output-generating atom $\phi_2$ in Corollary \ref{cor2}.
In Section \ref{example}, we show, by means of the analysis of example
signals, how the second layer output distinguishes tones that have a~smooth onset
(transient) from those that have a~sharp attack, which leads to broadband
characteristics of $A_n$ around this attack. Similarly, if $A_n$ undergoes an amplitude
modulation, the frequency of this modulation can be clearly discerned, cf. Figure
\ref{sharp_modul} and the corresponding example. This observation is clearly reflected
in expression Equation \eqref{eq:PropLayer2}.  
\end{Remark}

\begin{proof}
Using the outcome of Proposition \ref{prop}, we obtain
\begin{align}
&U_2[\beta_2 h] U_1[\beta_1j]f(m)=\notag \\ 
&|\langle S_{\alpha_1}(A_{n_0})  |\hat{g}_1(\beta_1 j -\xi_0 n_0)| +E_1, M_{\beta_2 h} T_{\alpha_2 m }g_2 \rangle_{\ell^2(\mathbb{Z})}|\leq \notag \\
&|\langle S_{\alpha_1}(A_{n_0})  |\hat{g}_1(\beta_1 j -\xi_0 n_0)| , M_{\beta_2 h} T_{\alpha_2 m }g_2 \rangle_{\ell^2(\mathbb{Z})}|  +  |\langle E_1, M_{\beta_2 h} T_{\alpha_2 m }g_2 \rangle_{\ell^2(\mathbb{Z})}|. \notag
\end{align}

For  the error $E_1(k)$, we use the global  estimate $ |\langle E_1, M_{\beta_2 h} T_{\alpha_2 m }g_2 \rangle_{\ell^2(\mathbb{Z})}| \leq \|E_1\|_\infty \cdot  \|g\|_1.$ Moreover, using the notation above and ignoring the constant term $|\hat{g}_1(\beta_1 j -\xi_0 n_0)|$, we proceed as follows,
\begin{equation}
\begin{array}{ll}
&\langle S_{\alpha_1}(A_{n_0}) , M_{\beta_2 h} T_{\alpha_2 m }g_2 \rangle_{\ell^2(\mathbb{Z})} =\sum_{k\in \mathbb{Z}} S_{\alpha_1}(A_{n_0}(k)) T_{\alpha_2 m} g_2( k) e^{-2\pi i \beta_2 h  k}=  \\ 
&\mathcal{F}\big{(} S_{\alpha_1}(A_{n_0}) \cdot T_{\alpha_2 m} g_2\big{)}({ \beta_2 h})= \mathcal{F}\big{(} S_{\alpha_1}(A_{n_0})\big{)} \ast \mathcal{F}\big{(}T_{\alpha_2 m} g_2\big{)}({ \beta_2 h})= \\
& P_{\frac{1}{\alpha_1}}\big{(} \hat{A}_{n_0} \big{)} \ast \big{(} M_{-\alpha_2 m} \hat{g}_2\big{)}({ \beta_2 h})=\bigg{(} \sum_{k \in \mathbb{Z}}\hat{A}_{n_0}\Big{(}.-\frac{k}{\alpha_1}\Big{)} \bigg{)} \ast \bigg{(} M_{-\alpha_2 m} \hat{g}_2\bigg{)}({ \beta_2 h}). \label{eq:period}
\end{array}
\end{equation} 

As $\hat{g}$ is concentrated around $0$, the right-hand term in Equation \eqref{eq:period} can only contain significant values if $A_{n_0}$ has frequency-components concentrated around $\beta_2 h,$ therefore we consider the case $k=0$ separately and obtain 
\begin{equation}
\begin{array}{ll}
\langle S_{\alpha_1}(A_{n_0}) , M_{\beta_2 h} T_{\alpha_2 m }g_2 \rangle_{\ell^2(\mathbb{Z})} &=\big{(} \hat{A}_{n_0}  \ast M_{-\alpha_2 m} \hat{g}_2\big{)}({ \beta_2 h })  \\
&+\left( \sum_{k \in \mathbb{Z}\setminus \{0\}}\hat{A}_{n_0}\Big{(}.-\frac{k}{\alpha_1}\Big{)} \right) \ast \bigg{(} M_{-\alpha_2 m} \hat{g}_2\bigg{)}({ \beta_2 h}). \label{eq:sumsplit}
\end{array}
\end{equation}

It remains to bound the sum of aliases, i.e., the second term of Equation \eqref{eq:sumsplit}:
\begin{equation}
\begin{array}{cl}
&\bigg{|}\bigg{(} \sum_{k \in \mathbb{Z}\setminus \{0\}}\hat{A}_{n_0}\Big{(}.-\frac{k}{\alpha_1}\Big{)} \bigg{)} \ast \bigg{(} M_{-\alpha_2 m} \hat{g}_2\bigg{)}({ \beta_2 h})\bigg{|}=  \\
&\bigg{|}\sum_r\bigg{(} \sum_{k \in \mathbb{Z}\setminus \{0\}}\hat{A}_{n_0}\Big(r-\frac{k}{\alpha_1}\Big) \bigg{)} \cdot \bigg{(} M_{-\alpha_2 m} \hat{g}_2\bigg{)}({ \beta_2 h-r})\bigg{|} \leq  \\
& \sum_r  \sum_{k \in \mathbb{Z}\setminus \{0\}} \bigg{|}\hat{A}_{n_0} \Big{(}r-\frac{k}{\alpha_1}\Big{)}\bigg{|}\cdot \bigg{|} \hat{g}_2(\beta_2h-r)\bigg{|} \label{eq:triangle}
\end{array}
\end{equation}

Using the assumption $ \sum_{k \in \mathbb{Z}\setminus \{0\}} | \hat{A}_{n_0}(.-\frac{k}{\alpha_1})| \leq \varepsilon_{\alpha_1}$ and also the assumption on the analysis window $g_2,$ namely, the fast decay of $\hat{g_2} $, we obtain
\begin{equation}
\begin{array}{ll}
\sum_r  \sum_{k \in \mathbb{Z}\setminus \{0\}} \bigg{|}\hat{A}_{n_0} \left(r-\frac{k}{\alpha_1}\right)\bigg{|}\cdot \bigg{|} \hat{g}_2(\beta_2h-r)\bigg{|} 
&\leq \varepsilon_{\alpha_1} \sum_r\big{|} \hat{g}_2(\beta_2h-r)\big{|} \\
&\leq \varepsilon_{\alpha_1} C_{\hat{g}_2} \sum_r \big{(} 1+|\beta_2h-r|^s\big{)}^{-1}.
\end{array}
\end{equation}

We rewrite the first term in  Equation \eqref{eq:sumsplit} and make use of the operator $\mathcal{I}$ introduced in Equation~\eqref{invol}:
\begin{equation}
\begin{array}{cl}
&\big{(} \hat{A}_{n_0}  \ast M_{-\alpha_2 m} \hat{g}_2\big{)}({ \beta_2 h })= \sum_r \hat{A}_{n_0}(r) \big{(}M_{-\alpha_2 m} \hat{g}_2\big{)}(\beta_2 h -r)= \\
&\langle \hat{A}_{n_0}, T_{\beta_2 h} \mathcal{I} M_{-\alpha_2 m} \hat{g}_2 \rangle= -\langle A_{n_0}, M_{\beta_2 h}  T_{\alpha_2 m} g_2 \rangle. \label{eq:planch}
\end{array}
\end{equation}

The last Equation \eqref{eq:planch} uses Plancherl's theorem. Rewriting the last term, we obtain 
$$-\langle A_{n_0}, M_{\beta_2 h}  T_{\alpha_2 m} g_2 \rangle= -\langle M_{-\beta_2 h}A_{n_0},   T_{\alpha_2 m} g_2 \rangle.$$
\begin{flushright}
\qedhere
\end{flushright}
\end{proof}
 
 \begin{Remark}
For sufficiently big $s$ the sum $\sum_r \big{(} 1+|\beta_2h-r|^s\big{)}^{-1}$ decreases  fast, e.g., taking $s = 5$  the sum is approximately $2.$

\end{Remark}
\noindent
The second layer output is obtained by applying the output-generating atom as in Equation \eqref{eq:output}.

\begin{Corollary}[Output of Layer $2$]
\label{cor2}
Let $\phi_2 \in \Psi_3,$ then the output of the second layer is 
\begin{align}
&\Big{(}U_2 [\beta_2 h]U_1 [\beta_1 j]f\ast \phi_2\Big{)}(m) =  \Big{(}|\hat{g}_1(\beta_1 j -\xi_0 n_0)| |\big{\langle}M_{-\beta_2 h} A_{n_0}, T_{\alpha_2 m} g_2\big{\rangle}|\ast \phi_2\Big{)}(m) +\epsilon_2(m)\notag
\end{align}
where
$$\epsilon_2(m)\leq \|E_2\|_\infty^2\|\phi_2\|_1^2.$$

Here $E_2$ is the error of Proposition \ref{cor}.
\end{Corollary}

\begin{Remark} 
Note that in the second layer, applying the output-generating atom $\phi_2 \in \Psi_3$ removes the fine temporal structure,
and thus the second layer output reveals information contained in the envelopes $A_n$.
\end{Remark}

\begin{proof}
Proof is similar to the first layer output, see Corollary \ref{Cor1}.
\begin{flushright}
\qedhere
\end{flushright}
\end{proof}

\subsubsection{Deformation Stability}
\label{deform}
\noindent
In this section, we study the extent  to which GS is stable with respect to
certain, small deformations. This question is interesting, as we may often
intuitively assume that the classification of natural signals, be it sound or images,
is preserved under mild and possibly local deformations. For the signal class
$\mathcal{T}$, we consider musically  meaningful deformations and show stability of
GS  with respect to these deformations.      We consider changes in
spectral shape as well as frequency modulations. Note that, as opposed to the
invariance properties derived in Section~\ref{scatmodel} for the output of specific layers, the derived stability results
pertain to the entire feature vector obtained from the GS along all included layers, cp. the definition and derivation of deformation stability in~\cite{mallat}. The method we apply is inspired by 
the authors of \cite{7541482} and uses the decoupling technique, i.e., to
prove stability of the feature extractor we first take the structural properties of the
signal class into account and search for an error bound of deformations of the signals
in  $\mathcal{T}.$ In combination with the contractivity property  
$\|\Phi_\Omega(f)-\Phi_\Omega(h)\|_2\leq \|f-h\|_2$  of $\Phi_\Omega$, see (\cite{DBLP:journals/corr/WiatowskiB15a}  Proposition 4), 
which follows from $B_\ell \leq 1 \,\,\,\forall \ell\in \mathbb{N},$ where $B_\ell$
is the upper frame bound of the Gabor frame $G(g_\ell, \alpha_\ell, \beta_\ell), $
this yields deformation stability of the feature~extractor.

Simply deforming a~tone would correspond to deformations of the envelope $A_n, \,\,n = 1,...,N.$ 
This corresponds to a~change in timbre, for example, by playing a~note on a~different
instrument.
Mathematically this can be expressed as
$\mathfrak{D}_{A_\tau} (f)(t)= \sum_{n=1}^N A_n\big{(}t+\tau(t)\big{)}e^{2 \pi i n \xi_0 t}.$
\begin{Lemma}[Envelope Changes]
\label{Lemma1}
Let $f \in \mathcal{T}$ and $|A'_n(t)| \leq C_n (1+|t|^s)^{-1}, $ for constants $C_n>0, \,\, n=1,...,N$ and $s>1.$ Moreover let $\| \tau \|_\infty < \frac{1}{2}.$ Then
$$
\|f- \mathfrak{D}_{A_\tau}(f)\|_2 \leq D \|\tau \|_\infty \sum_{n=1}^N C_n,\notag
$$
for $D>0$ depending only on $\|\tau\|_\infty.$
\end{Lemma}
 
\begin{proof}
Setting  $h_n(t) = A_n(t) -\mathfrak{D}_{A_\tau}(A_n(t)) $, we obtain
$$
\|f- \mathfrak{D}_{A_\tau}(f)\|_2 \leq \sum_{n=1}^N  \| h_n(t)\|_2.\notag
$$

We apply the mean value theorem for a~continuous function $A_n(t)$ and get
\begin{align}
|h_n(t)| \leq\| \tau \|_\infty \underset{y \in B_{\| \tau \|_\infty}(t)}{\mbox{sup}}  |A'_n(y)|.\notag
\end{align} 

Applying the $2-$norm on $h_n(t)$ and the assumption on $A_n'(t),$ we obtain

\begin{align}
\int_\mathbb{R}|h_n(t)|^2 dt &\leq \int_\mathbb{R}  \| \tau \|_\infty^2 \bigg{(} \underset{y \in B_{\| \tau \|_\infty}(t)}{\mbox{sup}}  |A'_n(y)|\bigg{)}^2 dt\notag \\
&\leq C_n^2  \| \tau \|_\infty^2 \int_{\mathbb{R}}\underset{y \in B_{\| \tau \|_\infty}(t)}{\mbox{sup}}  (1+|y|^s)^{-2} dt.\notag
\end{align}

Splitting the integral into $B_1(0)$ and $\mathbb{R}\backslash B_1(0)$,  we obtain
\begin{align}
\|h_n(t)\|_2^2 \leq C_n^2  \| \tau \|_\infty^2 \bigg{(} \int_{B_1(0)} 1 dt + \int_{\mathbb{R}\backslash B_1(0)} \underset{y \in B_{\| \tau \|_\infty}(t)}{\mbox{sup}}  (1+|y|^s)^{-2} dt\bigg{)}. \notag
\end{align}

Using the monotonicity of $(1+|y|^s)^{-1}$ and in order to remove the supremum, by shifting $ \| \tau \|_\infty,$  we have
\begin{align}
\|h_n(t)\|_2^2 \leq C_n^2  \| \tau \|_\infty^2 \bigg{(} \int_{B_1(0)} 1 dt + \int_{\mathbb{R}\backslash B_1(0)} (1+||t|-\|\tau\|_\infty|^s)^{-2}dt \bigg{)}. \notag
\end{align}

Moreover for $t \notin B_1(0)$  we have  $ |(1-\|\tau\|_\infty)t|^s \leq |(1-\frac{\|\tau\|_\infty}{|t|})t|^s.$
This leads to
$$
\|h_n(t)\|_2^2 \leq C_n^2  \| \tau \|_\infty^2 \bigg{(} 2 + \int_{\mathbb{R}\backslash B_1(0)}  (1+|(1-\| \tau \|_\infty)t|^s)^{-2}dt \bigg{)}.$$

Performing a~change of variables, i.e., $x = ( 1-  \| \tau \|_\infty) t$ with $\frac{dx}{dt} =1-\| \tau \|_\infty>\frac{1}{2}$ we obtain
\begin{align}
\|h_n(t)\|_2^2&\leq C_n^2  \| \tau \|_\infty^2 \bigg{(} 2+2 \int_{\mathbb{R}}(1+|x|^s)^{-2}dx\bigg{)}\notag\\
&=C_n^2  \| \tau \|_\infty^2 \bigg{(} 2+2 \bigg{\|} \frac{1}{1+|x|^s}\bigg{\|}^2_2\bigg{)}.\notag
\end{align}

Setting $D^2 := 2\big{(} 1+\| \frac{1}{1+|x|^s}\|^2_2\big{)}$ and summing up we obtain 
$$\|f- \mathfrak{D}_{A_\tau}(f)\|_2 \leq D \|\tau \|_\infty \sum_{n=1}^N C_n.$$
\begin{flushright}
\qedhere
\end{flushright}
\end{proof}

\begin{Remark}
Harmonics' energy decreases with increasing frequency, hence $C_n \ll C_{n-1},$ hence the sum $\sum_{n=1}^N C_n$ can be expected to be small. 
\end{Remark}
\noindent
Another kind of sound deformation results from frequency modulation of $f \in \mathcal{T}.$ This corresponds to, for example,  playing higher or lower pitch, or producing a~vibrato.
This can be formulated as
\begin{align}
\mathfrak{D}_{\tau}: f(t) \mapsto \sum_{n=1}^N A_n(t)e^{2 \pi i \big{(}n \xi_0 t+\tau_n(t)\big{)}}.\notag
\end{align}

\begin{Lemma}[Frequency Modulation]
\label{Lemma2}
Let $f \in \mathcal{T}.$ Moreover let $\| \tau_n \|_\infty< \frac{\arccos(1-\frac{\varepsilon^2}{2})}{2\pi}.$
Then,
\begin{align}
 \|f- \mathfrak{D}_{\tau}(f)\|_2 \leq   \varepsilon \sum_{n=1}^N \frac{1}{n}.\notag
\end{align}
\end{Lemma}

\begin{proof}
We have
$$
\|f-\mathfrak{D}_{\tau}f\|_2  \leq \sum_{n=1}^N \| h_n(t)\|_2, \notag
$$
with $h_n(t) = A_n(t) (1-e^{2 \pi i \tau_n(t)})$.
Computing the $2-$norm of $h_n(t)$, we obtain
\begin{align}
\int_\mathbb{R} |h_n(t)|^2dt &= \int_{\mathbb{R}} |A_n(t) (1-e^{2 \pi i \tau_n(t)})|^2dt \leq  \|1-e^{2 \pi i \tau_n(t)}\|^2_\infty \|A_n(t)\|^2_\infty. \notag
\end{align}

We rewrite
 $$|1-e^{2 \pi i \tau_n(t)}|^2 = \big{|}1-\big{(}\cos\big{(}2 \pi \tau_n(t)\big{)}+i\sin\big{(}2 \pi  \tau_n(t)\big{)}\big{)}\big{|}^2 = 2\left(1-\cos\big{(}2\pi \tau_n(t)\big{)}\right).$$
 
 Setting $\|1-e^{2 \pi i \tau_n(t)}\|^2_\infty \leq \varepsilon^2,$ this term gets small if $\|\tau_n(t)\|_\infty \leq\frac{\arccos(1-\frac{\varepsilon^2}{2})}{2\pi}.$
Using the assumptions of our signal model on the envelopes, i.e., $\|A_n\|_\infty < \frac{1}{n},$ we obtain
$$\|f- \mathfrak{D}_{\tau}(f)\|_2 \leq \varepsilon \sum_{n=1}^N \frac{1}{n}.$$
\begin{flushright}
\qedhere
\end{flushright}
\end{proof}

\begin{Proposition}[Deformation Stability]
Let  $\Phi_\Omega: L^2(\mathbb{R}) \to ( \ell^2(\mathbb{Z}))^\mathcal{Q},$ $f \in \mathcal{T}$ and $|A'_n(t)| \leq C_n (1+|t|^s)^{-1}, $ for constants $C_n>0, \,\, n=1,...,N$ and $s>1.$ Moreover let $\| \tau \|_\infty < \frac{1}{2}$ and $\| \tau_n \|_\infty< \frac{\arccos(1-\frac{\varepsilon^2}{2})}{2\pi}.$
Then the feature extractor $\Phi$ is deformation stable with respect to
\begin{itemize}
    \item envelope changes $\mathfrak{D}_{A_\tau} \big{(}f\big{)}(t)= \sum_{n=1}^N A_n\big{(}t+\tau(t)\big{)}e^{2 \pi i n \xi_0 t}:$
    $$\big{\|}\Phi_\Omega\big{(}f\big{)}-\Phi_\Omega\big{(}\mathfrak{D}_{A_\tau}(f)\big{)}\big{\|}_2\leq  D \|\tau \|_\infty \sum_{n=1}^N C_n,$$
    for $D>0$ depending only on $\|\tau\|_\infty.$
    \item frequency modulation $\mathfrak{D}_{\tau}\big{(}f\big{)}(t)= \sum_{n=1}^N A_n(t)e^{2 \pi i \big{(}n \xi_0 t+\tau_n(t)\big{)}}: $
    $$\big{\|}\Phi_\Omega\big{(}f\big{)}-\Phi_\Omega\big{(}\mathfrak{D}_{\tau}(f)\big{)}\big{\|}_2\leq
    \varepsilon \sum_{n=1}^N \frac{1}{n}.$$
\end{itemize}
\end{Proposition}

\begin{proof}
The Proof follows directly from a~result of (\cite{DBLP:journals/corr/WiatowskiB15a}  Proposition 4), called contractivity property  
$\|\Phi_\Omega(f)-\Phi_\Omega(h)\|_2\leq \|f-h\|_2$  of $\Phi_\Omega,$ which follows from $B_\ell \leq 1 \,\,\,\forall \ell\in \mathbb{N},$ where $B_\ell$ is the upper frame bound of the Gabor frame $G(g_\ell, \alpha_\ell, \beta_\ell) $ and deformation stability of the signal class in Lemmas~\ref{Lemma1}~and \ref{Lemma2}. 
\begin{flushright}
\qedhere
\end{flushright}
\end{proof}

\subsection{Visualization Example} \label{example}
\noindent
 
\noindent
In this section, we present some visualizations based on two implementations, one in \textsc{Matlab}, which we call the GS implementation, and the other one in Python, which is the channel averaged GS implementation. The main difference between these implementations is an averaging step of Layer~2 in the case of the Python implementation; averaging over channels is introduced in order to obtain a~ 2D representation  in each layer. Furthermore, the averaging step significantly accelerates the computation of  the second layer output.

\begin{figure}
\centering
  \includegraphics[width=4in]{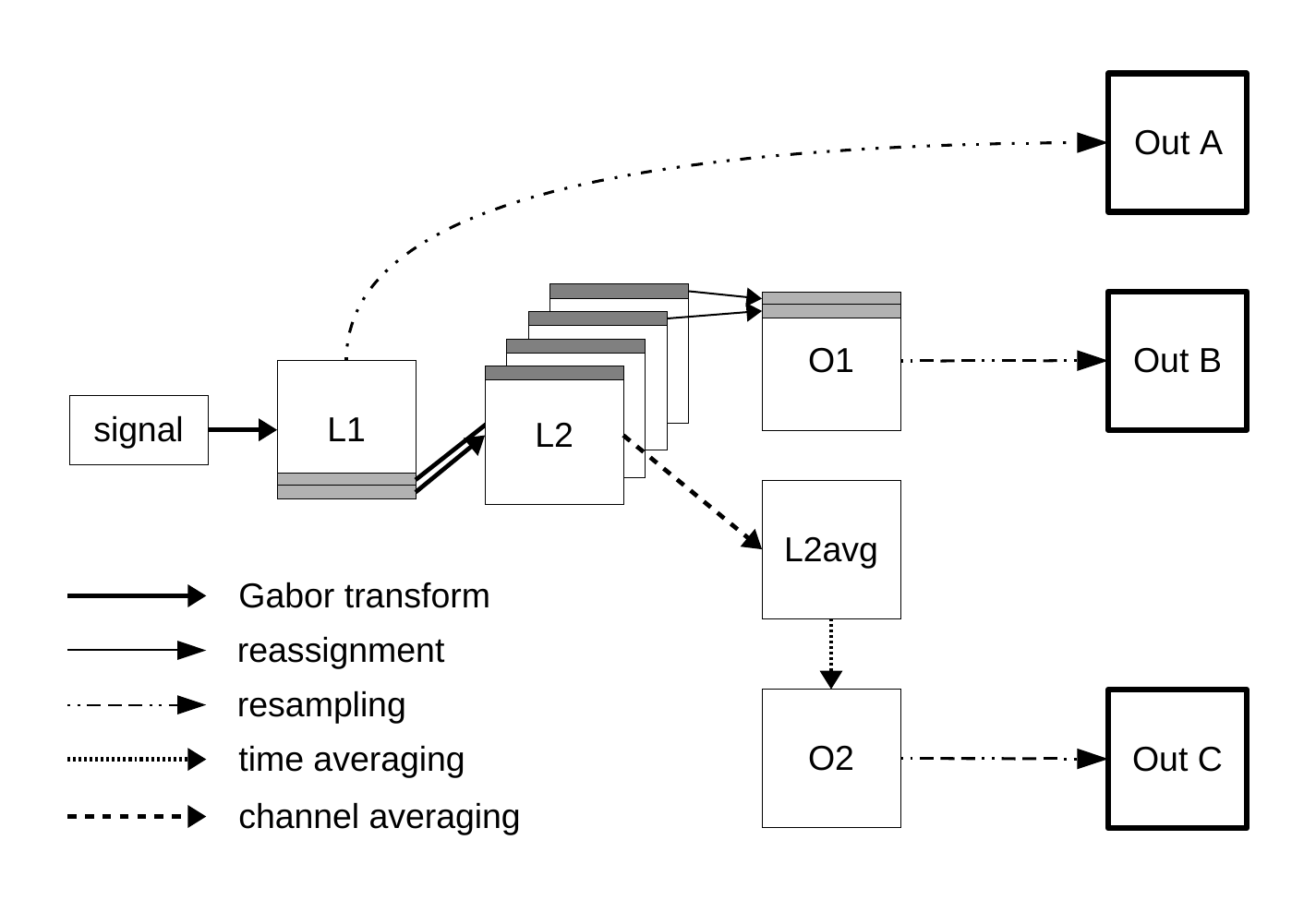}
\caption{Diagram explaining the naming of the GS building blocks of the Python implementation in the following sections.} 
\label{fig:diagram}
\end{figure}

Referring to Figure \ref{fig:diagram}, the following nomenclature will be used. Layer~1~(L1) is the GT, which, after resampling to the desired size, becomes Out~A. Output~1~(O1) is the output of L1, i.e., after applying the output-generating atom. Recall that this is done by a~low-pass filtering step. Again, Out~B  is obtained by  resampling to the desired matrix size.

Layer~2~(L2) is obtained by applying another GT for each frequency channel. In the \textsc{Matlab} code,  Output~2 (O2) is then obtained by low-pass filtering the separate channels of each resulting spectrogram. In the case of Python implementation (see Figure \ref{fig:diagram}), we average all the GT of L2 to one spectrogram (for the sake of speed) and then apply a~time averaging step in order to obtain O2. Resampling to the desired size yields Out~C.

As input signal for this section we generate single tones following the signal model from Section~\ref{model}.

\subsubsection{Visualization of Different Frequency Channels within the GS Implementation}
\begin{figure}[H]
\centering
   \includegraphics[width=5.5in]{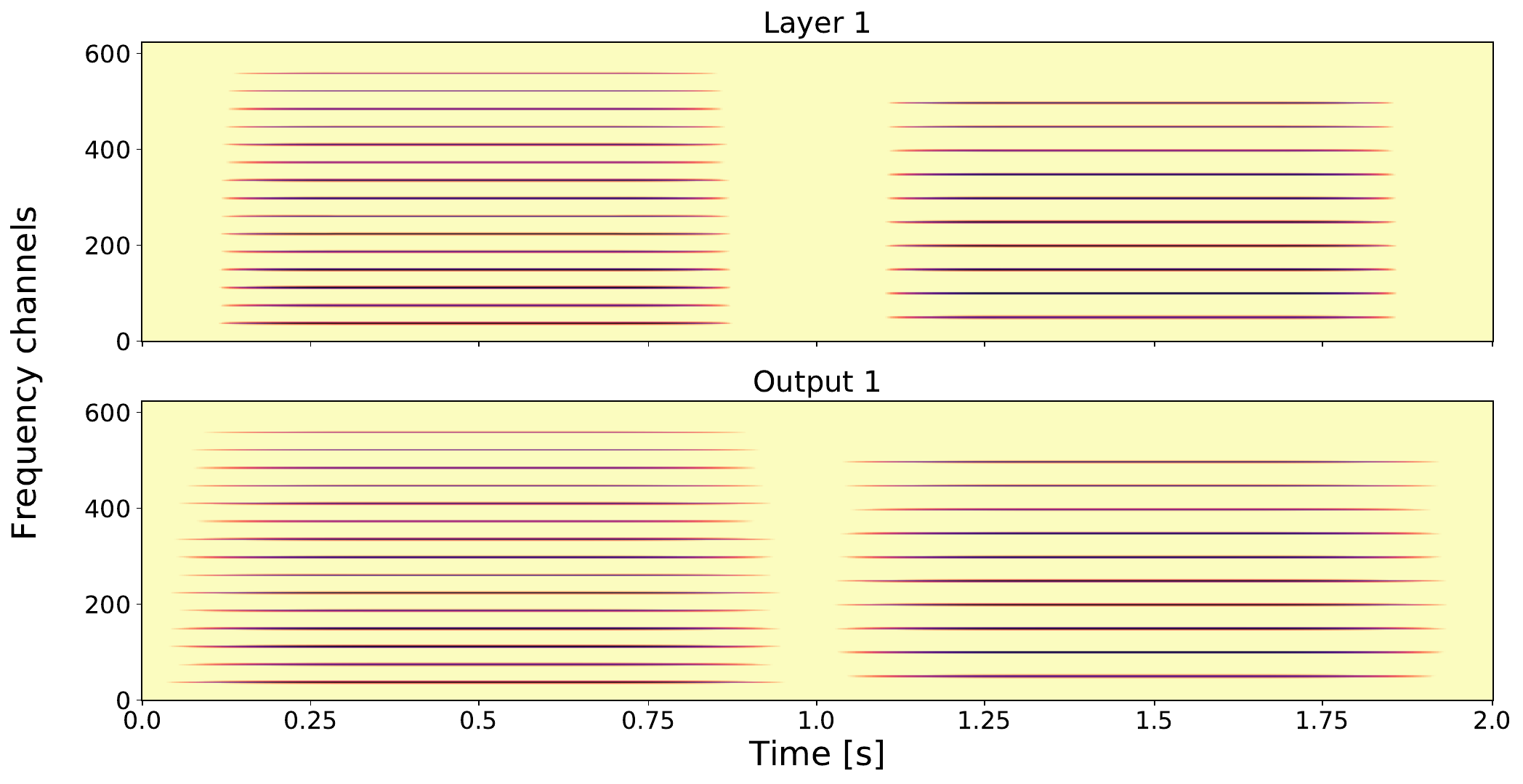}
\caption{First layer (i.e., GT) and Output~$1$ of two tones with different fundamental frequencies.} 
\label{2tones}
\end{figure}

\begin{figure}[H]
\centering
   \includegraphics[width=4.5in]{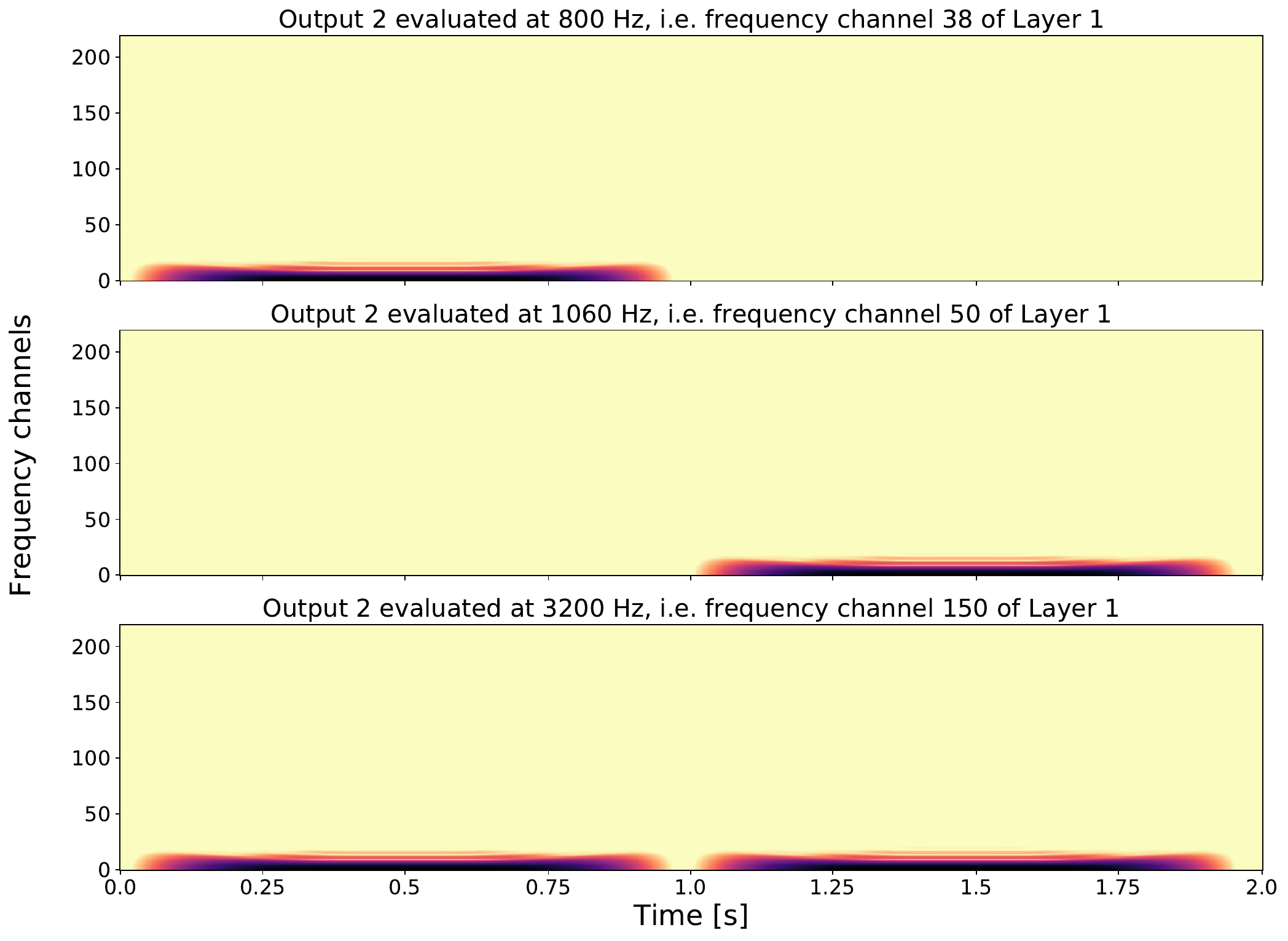}
\caption{Output~$2$ of two tones with different fundamental frequencies, at different fixed frequency channels of Layer~1.} 
\label{different_pitch}
\end{figure}

\noindent
Figures \ref{2tones} and  \ref{different_pitch} show two tones, both having a~smooth envelope but different fundamental frequencies and number of harmonics. The first tone has fundamental frequency $\xi_0 = 800\,\mbox{Hz}$ and $15$~harmonics, and the second tone has fundamental frequency $\xi_0 = 1060\,\mbox{Hz}$ and $10$  harmonics.

\noindent
Content of Figures  \ref{2tones} and  \ref{different_pitch}:
\begin{itemize}
\item \textit{Layer~1:} The first spectrogram of Figure \ref{2tones} shows the GT. Observe the difference in the fundamental frequencies and that these two tones have a~different number of harmonics, i.e., tone one has more than tone two.
\item \textit{Output~1:} The second spectrogram of Figure \ref{2tones} shows Output~1, which is is time averaged version of Layer~1.
 \item \textit{Output~2:}
For the second layer output (see Figure \ref{different_pitch}),  we take a~fixed frequency channel from Layer~1 and compute another GT to obtain a~Layer~2 element. By applying an output-generating atom, i.e., a~low-pass filter,  we obtain Output~2. Here, we show how different frequency channels of Layer~1 can affect Output~2.
The first spectrogram shows Output 2 with respect to, the fundamental frequency of tone one, i.e., $\xi_0 = 800\,\mbox{Hz}.$ Therefore no second tone is visible in this output. On the other hand, in the second spectrogram, if  we take as fixed frequency channel in Layer~1 the fundamental frequency of the second tone, i.e., $\xi_0 = 1060\,\mbox{Hz},$ in Output~2, the first tone is not visible. 
If we consider a~frequency that both share, i.e., $\xi = 3200\,\mbox{Hz}$, we see that for Output~2 in the third spectrogram both tones are present.  As GS focuses on one frequency channel in each layer element, the frequency information in this layer is lost; in other words, Layer~2 is invariant with respect to frequency.
\end{itemize}

\subsubsection{Visualization of Different Envelopes within the GS Implementation}

\begin{figure}[H]
\centering
  \includegraphics[width=5.5in]{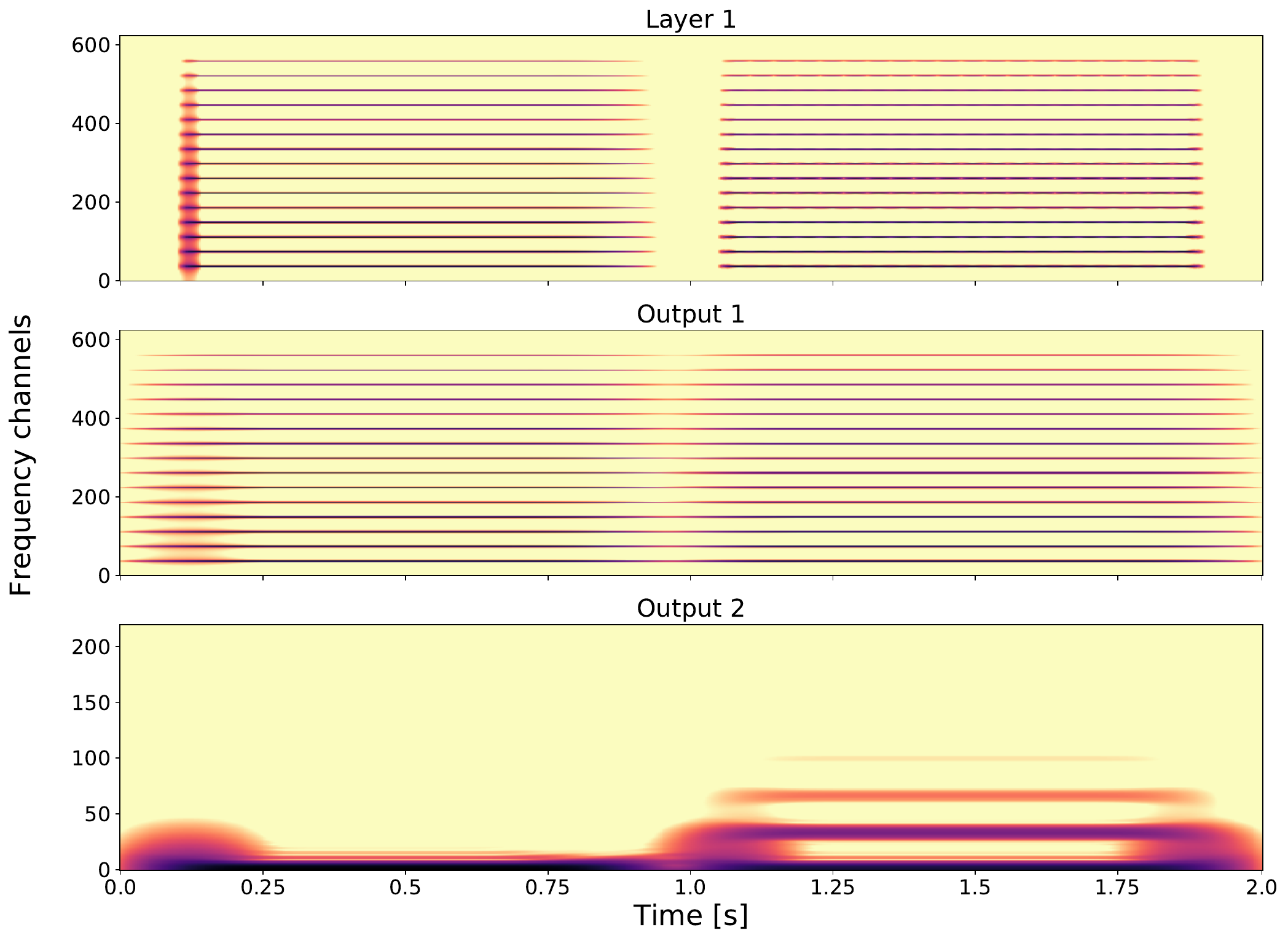}
\caption{Layer~$1$ (i.e., GT), Output~$1$, and Output~$2$ of the signal having a~sharp attack and afterwards some modulation.} 
\label{sharp_modul}
\end{figure}

\noindent
Here, Figure \ref{sharp_modul} shows two tones, played sequentially, having the same fundamental frequency $\xi_0 = 800\,\mbox{Hz}$ and $15$~harmonics, but different envelopes. The first tone has a~sharp attack, maintains and goes softly to zero, the second starts with a~soft attack and has some amplitude modulation. An~amplitude modulated signal would, for example, correspond to $f(t) = \sum_{n=1}^N \sin(2\pi 20t)e^{2\pi i n \xi_0 t}$; here, the signal is modulated by $20Hz.$ The GS output of these signals are shown in Figure \ref{sharp_modul}. \\

\begin{itemize}
\item \textit{Layer~1:} In the spectrogram showing the GT, we see the difference between the envelopes and we see that the signals have the same pitch and the same harmonics.
\item \textit{Output~1:} {The output of the first layer is invariant with respect to the envelope of the signals. This is due to the output-generating atom and the subsampling, which removes temporal information of the envelope.  In this output, no information about the envelope (neither the sharp attack nor the amplitude modulation) is visible, therefore the spectrogram of the different signals look almost the~same.}
\item \textit{Output~2:} For the second layer output we took as input a~time vector at fixed frequency of 800 Hz (i.e., frequency channel~38) of the first layer. Output~2 is invariant with respect to the pitch, but differences on larger scales are captured. Within this layer we are able to distinguish the different envelopes of the signals. We first see the sharp attack of the first tone and then the modulation with a~second frequency is visible.
\end{itemize}
\noindent

The source code of the \textsc{Matlab} implementation and further examples can be found in~\cite{gs-gt}. 

\subsubsection{Visualization of How Frequency and Amplitude Modulations Influence the Outputs Using the Channel Averaged Implementation}
\begin{figure}[H]
\centering
   \includegraphics[width=3.3in]{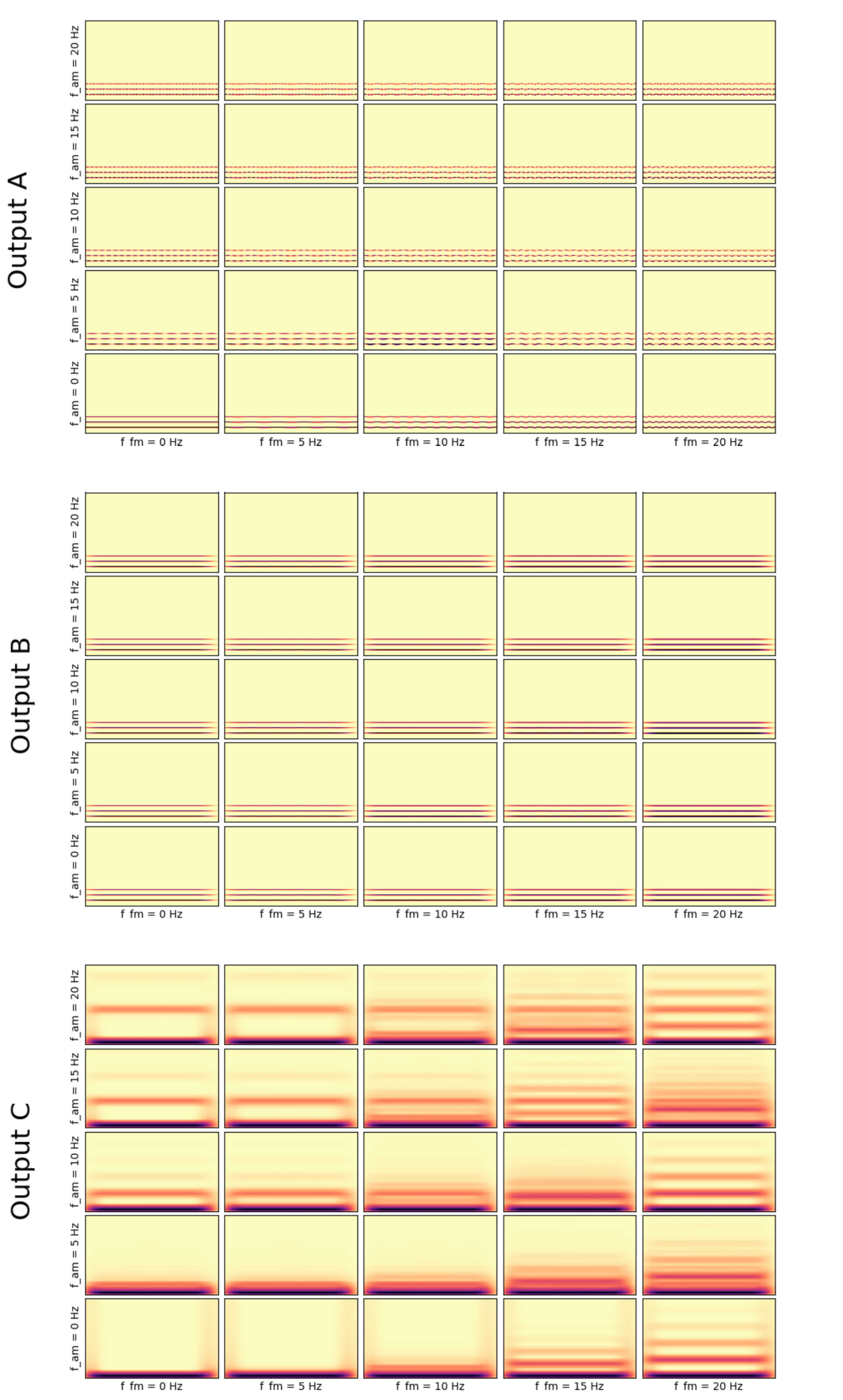}
\caption{Example output of the interactive plot.} 
\label{grid}
\end{figure}

\noindent
To visualize the resampled transformation in a~more structured way, we created an interactive plot (see Figure \ref{grid}), which shows 25~different synthetic audio signals side by side, transformed into Out~A, Out~B, and Out~C with chosen GS parameters. Each signal consists of one or more sine waves modulated in amplitude and frequency with 5\,Hz steps.

\noindent
The parameters can be adjusted by sliders and the plot is changed accordingly. The chosen parameters to be adjusted were number of frequency channels in Layer~1, number of frequency channels in Layer~2, sampling rate, and number of harmonics of the signal. 
The code for the interactive plot is available as a~part of the repository~\cite{gs-gt}.

\section{Experimental Results}
\label{section:expresults}
\noindent
In the numerical experiments, we compare the GS to a~GT representation, which is one of the standard time-frequency representations used in a~preprocessing phase for training neural networks applied to audio data. We compare these two time-frequency representations with respect to the performance on limited size of the training dataset.

To convert the raw waveform into the desired representations (GS and GT), we have used the Gabor-scattering~v0.0.4 library~\cite{gabor-scattering}, which is our Python implementation of the GS transform based on the Scipy~v1.2.1~\cite{scipy, oppenheim1999discrete, griffin1984signal} implementation of STFT.

{To demonstrate the beneficial properties of GS, we first create synthetic data in which we have the data generation under a~full control. In this case, we generate four classes of data that reflect the discriminating properties of GS. Second, we investigate whether the GS representation is beneficial when using a~``real'' dataset for training. For this purpose, we have utilized the GoodSounds dataset~\cite{GoodSounds}}.

\subsection{Experiments with Synthetic Data}
\noindent
In the synthetic dataset, we created four classes containing $1$\,s long signals, sampled at $44.1$\,kHz with $16$\,bit precision. All signals consist of a~fundamental sine wave and four harmonics. The whole process of generating sounds is controlled by fixed random seeds for reproducibility.
\subsubsection{Data}
\label{section:synthetic-data}
\noindent
We describe the sound generator model for one component of the final signal by the following~equation,
\begin{align}
    f(t) = A \cdot \sin\left(2\pi \big{(}\xi t + cw_{fm}(t, A_{fm}, \xi_{fm}, \varphi_{fm})\big{)} + \varphi\right) \cdot cw_{am}(t, A_{am}, \xi_{am}, \varphi_{am}),
\end{align}
where 
$cw_{fm}(t, A_{fm}, \xi_{fm}, \varphi_{fm})~=~A_{fm} \cdot \sin(2\pi \xi_{fm} t + \varphi_{fm})$ is the frequency modulation and
$cw_{am}(t, A_{am}, \xi_{am}, \varphi_{am})~=~\left\{\begin{array}{lr}
        A_{am} \cdot \sin(2\pi \xi_{am} t + \varphi_{am}) & \,\,\,\,\,\mathrm{  if }\ A_{am} > 0 \mbox{ and } \big{(}\varphi_{am}>0 \mbox{ or } \xi_{am}>0\big{)}\\
        
        1                                      & \mathrm{else}
        \end{array}\right\}
\vspace{15pt}$ is the amplitude modulation.
Here, $A$ is the amplitude, $\xi$ denotes the frequency and $\varphi$ denotes the phase. 
Furthermore, the amplitude, frequency, and phase of the frequency modulation carrier wave is denoted by $A_{fm}$, $\xi_{fm}$, and $\varphi_{fm}$, respectively, and for the case of amplitude modulation carrier wave we have $A_{am}$, $\xi_{am}$, and $\varphi_{am}.$

\noindent
To generate five component waves using the sound generator described above, we needed to decide upon the parameters of each component wave. We started by randomly generating
the frequencies and phases of the signal and the carrier waves for frequency and amplitude modulation
from given intervals. These parameters describe the fundamental sine wave of the signal. Next we create harmonics by taking multiples (from~2 to~5) of the fundamental frequency $\xi,$ where $A$ of each next harmonic is divided by a~factor. 
Afterwards, by permuting the two parameters, namely, by turning the amplitude modulation and frequency modulation on and off, we defined four classes of sound. These classes are indexed starting from zero. The $\mathrm{0}^{th}$ class has neither amplitude nor frequency modulation. Class~1 is just amplitude modulated, Class~2 is just modulated in frequency, and Class~3 is modulated in both amplitude and frequency, as seen in Table~\ref{table:1}. At the end, we used those parameters to generate each harmonic separately and then summed them together to obtain the final audio file. 

\begin{table}[H]
\centering
\caption{Overview of classes.}
\begin{tabular}{ccc} 
    \toprule 
    & $A_{am}$~=~0 & $A_{am}$~=~1 \\ 
    \midrule 
    $A_{fm}~=~0$ & class 0 & class 1 \\ 
    $A_{fm}$~=~1 & class 2 & class 3 \\ 
    \bottomrule 
\end{tabular}
\label{table:1}
\end{table}
\noindent
The following parameters were used to obtain GS; n\_fft\,=\,500---number of frequency channels, n\_perseg\,=\,500---window length, n\_overlap =\,250---window overlap were taken for Layer~1, i.e., GT, n\_fft\,=\,50, n\_perseg\,=\,50, n\_overlap\,=\,40 for Layer~2, window\_length of the time averaging window for Output~2 was set to~5 with mode set to ``same''. All the shapes for Output~A, Output~B, and Output~C were $240 \times 160$. Bilinear resampling~\cite{bilinear} was used to adjust the shape if necessary. The same shape of all of the outputs allows the stacking of matrices into shape $3\times240\times160$, which is convenient for CNN, because it can be treated as a~3-channel image. Illustration of the generated sounds from all four classes transformed into GT and GS can be seen in Section \ref{inv_gs} and Figure~\ref{fig:visualization_synth}. 

With the aforementioned parameters, the mean time necessary to compute the GS was 17.4890\,ms, whereas the mean time necessary to compute the GT was~5.2245\,ms, which is approximately $3$~times less. Note that such comparison is only indicative, because the time is highly dependent on chosen parameters, hence the final time depends on the specific settings.

\subsubsection{Training}
\label{sec:synthTraining}
\noindent
To compare the discriminating power of both GS and GT, we have generated 10,000~training samples ($2500$ from each class) and 20,000 ($5000$ from each class) validation samples. As the task at hand is not as challenging as some real-world datasets, we assume these sizes to be sufficient for both time-frequency representations to converge to very good performances. 
To compare the performance of GS and GT on a~limited set of training data, we have altogether created four scenarios in which the training set was limited to $400, 1000, 4000$, \mbox{ and } 10,000~samples. In all of these scenarios, the size of the validation set remained at its original size of 20,000~samples and we have split the training set into smaller batches each containing~100 samples with the same number of samples from each class. Batches were used to calculate the model error based on which the model weights were updated.

\noindent
The CNN consisted of the batch normalization layer, which acted upon the input data separately for each channel of the image (we have~three channels, namely Out~A, Out~B, and Out~C), followed by four stacks of~2D convolution with average pooling. The first three convolutional layers were identical in the number of kernels, which was set to~16 of the size $3\times3$ with stride $1\times1$. The last convolutional layer was also identical apart from using just~8 kernels. Each convolutional layer was initialized by a Glorot uniform initialization~\cite{glorot}, and followed by a~ReLu nonlinearity~\cite{relu} and an average pooling layer with a~$2\times2$ pool size. After the last average pooling the feature maps  were flattened and fully connected to an output layer with $4$ neurons and a~softmax activation function~\cite{softmax}. For more details about the networks architecture, the reader should consult the repository~\cite{gs-gt}. There one also finds the exact code in order to reproduce the experiment. 

\noindent
The network's categorical cross-entropy loss function was optimized using the Adam optimizer~\cite{adam} with $lr$ = 0.001, $\beta_1$ = 0.9, and $\beta_2$ = 0.999. To have a fair comparison, we limit each of the experiments in terms of computational effort as measured by a~number of weight updates during the training phase. One weight update is made after each batch. Each experiment with synthetic data was limited to~2000 weight updates. To create the network, we used Python~3.6~programming language with Keras framework~v2.2.4~\cite{chollet2015keras} on Tensorflow backend~v1.12.0~\cite{tensorflow2015-whitepaper}. To train the models, we used two GPUs, namely, NVIDIA Titan~XP and NVIDIA GeForce~GTX~1080~Ti, on the OS~Ubuntu~18.04 based system. Experiments are fully reproducible and can be obtained by running the code in the repository~\cite{gs-gt}.

\subsubsection{Results}

\noindent
The results are shown in Table~\ref{table:synthresults}, listing the accuracies of the model's best weight update on training and validation sets. The best weight update was chosen based on the performance on the validation set. More detailed tables of the results can be found in the aforementioned repository. In~this experiment, we did not use any testing set, because of the synthetic nature of the data. Accuracy is computed as a~fraction of correct predictions to all predictions. 
\noindent

\begin{figure}[H]
\centering
\includegraphics[width=4.5in]{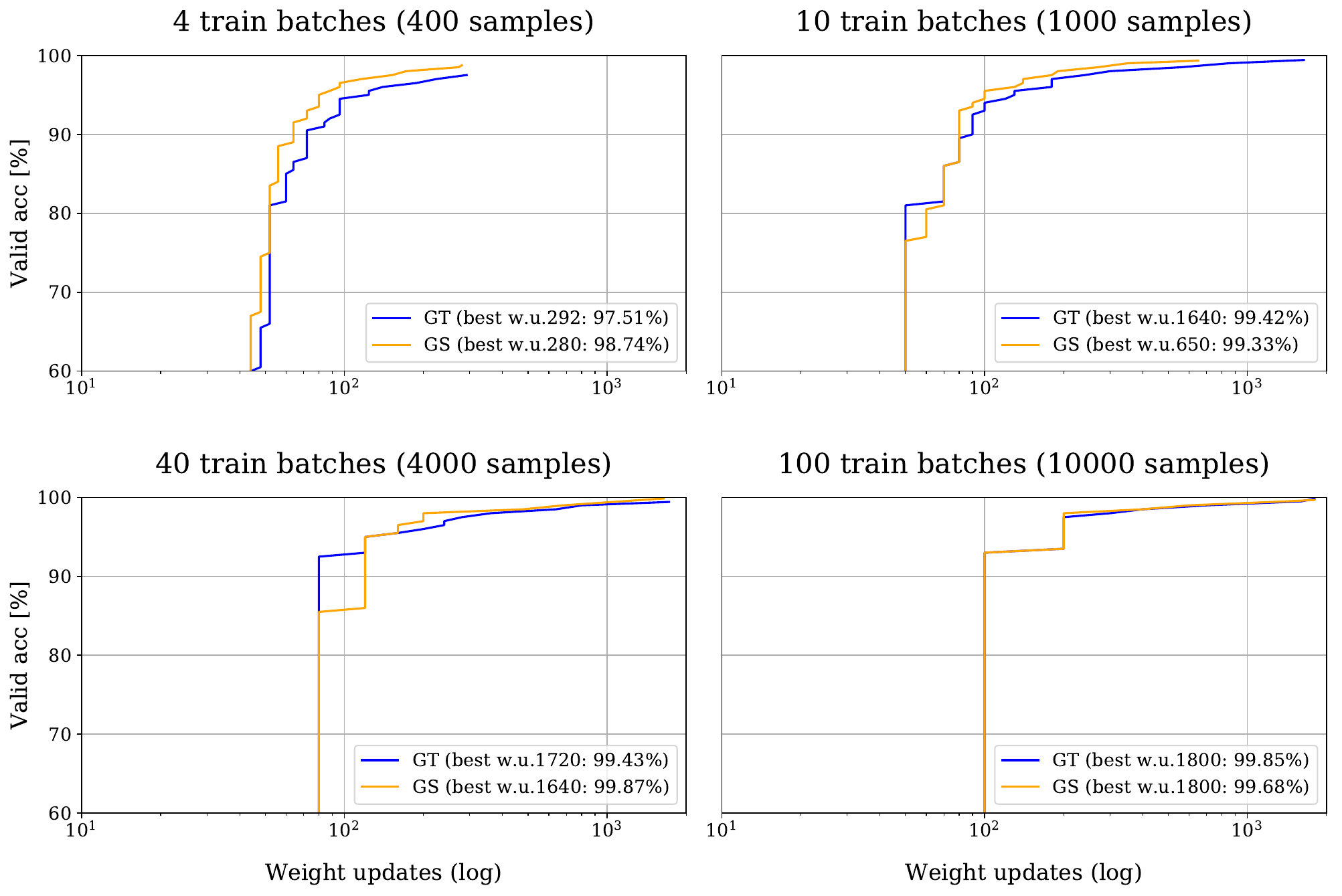}
\caption{CNN performance milestone reached over number of weight updates---Synthetic data. Figure~notation: 
Valid acc---Accuracy performance metric measured on the validation set. 
Best~w.u.---Weight update after which the highest performance was reached.}
\label{fig:performance}
\vspace{5mm}
\end{figure}

The most important observation is visible in Figure \ref{fig:performance}, where it is shown that in the earlier phases of the training, GS reaches higher accuracies after less weight updates than GT. This effect diminishes with bigger training sets and vanishes completely in case of~100 training batches. In case of very limited data, i.e., with only~400 training samples, the results show that GS even outperformed GT. With~more training samples, i.e.,~1000 and~4000, the best performances of GT and GS are nearly the same. In this case we could hypothesize that the prior knowledge of the intrinsic properties of a~time series signal shown by GS (in the invariances of Layer~1 and Layer~2) is not needed anymore and the network is able to learn the necessary transformation itself.

\begin{table}[H]
\centering
\caption{Performance of the convolutional neural network (CNN) trained using GS and GT data.}
\begin{tabular}{cccccc} 
\toprule 
\textbf{TF}	&	\textbf{N Train}	&	\textbf{N Valid}	&	\textbf{BWU}	&	\textbf{Train}	&	\textbf{Valid}	\\
\midrule 
GS	&	400	    &	20,000	&	280	&	1.0000	&	0.9874  \\
GT	&	400	    &	20,000	&	292	&	1.0000	&	0.9751	\\
\midrule
GS	&	1000	&	20,000	&	650	&	0.9990	&	0.9933	\\
GT	&	1000	&	20,000	&	1640	&	1.0000	&	0.9942	\\
\midrule
GS	&	4000	&	20,000	&	1640	&	0.9995	&	0.9987	\\
GT	&	4000	&	20,000	&	1720	&	0.9980	&	0.9943	\\
\midrule
GS	&	10,000	&	20,000	&	1800	&	0.9981	&	0.9968	\\
GT	&	10,000	&	20,000	&	1800	&	0.9994	&	0.9985	\\
\bottomrule 
\end{tabular}
\begin{tabular}{lrrrrr}
\multicolumn{1}{p{\textwidth -.88in}}{\footnotesize Table notation: TF---Time-frequency representation. 
N train and N valid---Number of samples in training and validation sets. BWU---Weight update after which the highest performance was achieved on the validation set. Train and valid---accuracy on training and validation sets.}
\end{tabular}
\label{table:synthresults}

\vspace{-12pt}  
\end{table}

\subsection{Experiments with GoodSounds Data}
\noindent
In the second set of experiments, we used the GoodSounds dataset~\cite{GoodSounds}. It contains monophonic audio recordings of single tones or scales played by~12 different musical instruments. The main purpose of this second set of experiments is to investigate whether GS shows superior performance to GT in a~classification task using real-life data.

\subsubsection{Data}
\noindent
To transform the data into desired form for training, we removed the silent parts using the SoX v14.4.2 library~\cite{sox, SoxSilenceTut}; next, we split all files into~1\,s long segments sampled at a~rate of~44.1\,kHz with~16\,bit precision. A~Tukey window was applied to all segments to smooth the onset and the offset of each with the aim to prevent undesired artifacts after applying the STFT.

\noindent
The dataset contains $28.55$\,h of recordings, which is a~reasonable amount of audio data to be used in training of Deep Neural Networks considering the nature of this task. Unfortunately, the data are distributed into classes unevenly, half of the classes are extremely underrepresented, i.e., half of the classes together contain only~12.6\% of all the data. In order to alleviate this problem, we decided upon an equalization strategy by variable stride. 

To avoid extensive equalization techniques, we have discarded all classes that spanned less than~10\% of the data. In total we used~six classes, namely, clarinet, flute, trumpet, violin, sax alto, and cello. To equalize the number of segments between these classes, we introduced the aforementioned variable stride when creating the segments. The less data a~particular class contains, the bigger is the overlap between segments, thus more segments are generated and vice versa. The whole process of generating sounds is controlled by fixed random seeds for reproducibility. Detailed information about the available and used data, stride settings for each class, obtained number of segments and their split can be seen in Table~\ref{table:goodsoundsdata}. 

\begin{table}[H]
\caption{Overview of available and used data.}
\label{table:goodsoundsdata}
\centering

\begin{tabular}{ccccccccc}
\toprule
\multicolumn{1}{c}{} &  \multicolumn{1}{c}{} & \multicolumn{3}{c}{\textbf{All Available Data}} & \multicolumn{4}{c}{\textbf{Obtained Segments}}\\
\midrule
& \textbf{Class}	 & 	\textbf{Files}	 & 	\textbf{Dur}	 & 	\textbf{Ratio} &  \textbf{Stride}	 & 	\textbf{Train} & 	\textbf{Valid}  & 	\textbf{Test} \\
\midrule
\multirow{6}{*}{Used} 
& Clarinet	 & 	3358	 & 	369.70	 & 	21.58\%    & 37,988  & 12,134     & 4000      & 4000  \\ 
& Flute	     & 	2308	 & 	299.00	 & 	17.45\%    & 27,412  & 11,796     & 4000      & 4000  \\ 
& Trumpet	 & 	1883	 & 	228.76	 & 	13.35\%    & 22,826  & 11,786     & 4000      & 4000  \\ 
& Violin	 & 	1852	 & 	204.34	 & 	11.93\%    & 19,836  & 11,707     & 4000      & 4000  \\ 
& Sax alto   & 	1436	 & 	201.20	 & 	11.74\%    & 19,464  & 11,689     & 4000      & 4000  \\ 
& Cello	     & 	2118	 & 	194.38	 & 	11.35\%    & 15,983  & 11,551     & 4000      & 4000  \\ 
\midrule
\multirow{6}{*}{Not used} 
& Sax tenor	 & 	680	    & 	63.00	 & 	3.68\%     &        & 		    & 		    &       \\ 
& Sax soprano & 668	    & 	50.56	 & 	2.95\%     &        & 		    & 		    &       \\ 
& Sax baritone & 576    & 	41.70	 & 	2.43\%     &        & 		    & 		    &       \\ 
& Piccolo	 & 	776	    & 	35.02	 & 	2.04\%     &        & 		    & 		    &       \\ 
& Oboe	     & 	494	    & 	19.06	 & 	1.11\%     &        & 		    & 		    &       \\ 
& Bass	     & 	159	    & 	6.53	 & 	0.38\%     &        & 		    & 		    & 		\\ 
\midrule
& Total	     & 	16,308	& 	1713.23	 & 	100.00\%   &        & 70,663    & 	24,000	& 24,000 \\ 
\bottomrule
\end{tabular}
\begin{tabular}{@{}c@{}} 
\multicolumn{1}{p{\textwidth -.88in}}{\footnotesize Table notation: 
Files---Number of available audio files. 
Dur---Duration of all recordings within one class in minutes.
Ratio---Ratio of the duration to total duration of all recordings in the dataset.
Stride---Step size (in~samples) used to obtain segments of the same length.
Train, Valid, Test---Number of segments used to train (excluding the leaking segments), validate, and test the model.}
\end{tabular}
 
\end{table}

As seen from the table, the testing and validation sets were of the same size comprising the same number of samples from each class. The remaining samples were used for training. To~prevent leaking of information from validation and testing sets into the training set, we have excluded all the training segments originating from the audio excerpts, which were already used in validation or testing set. More information can be found in the repository~\cite{gs-gt}.

The following parameters were used to obtain GS; n\_fft\,=\,2000---number of frequency channels, n\_perseg\,=\,2000---window length, n\_overlap\,=\,1750---window overlap were taken for Layer~1, i.e., GT, n\_fft\,=\,25, n\_perseg\,=\,25, n\_overlap\,=\,20 for Layer~2, window\_length of the time averaging window for Output~2 was set to~5 with mode set to `same'. All the shapes for Output~A, Output~B and Output~C were $480 \times 160$. Bilinear resampling~\cite{bilinear} was used to adjust the shape if necessary. The same shape of all the outputs allows the stacking of matrices into shape $3\times480\times160$. Illustration of the sounds from all six classes of musical instruments transformed into GT and GS can be found in the repository~\cite{gs-gt}.

\subsubsection{Training}
\noindent
To make the experiments on synthetic data and the experiments on GoodSounds data comparable, we again used the CNN as a~classifier trained in a similar way as described in Section~\ref{sec:synthTraining}. We have also preprocessed the data, so the audio segments are of the same duration and sampling frequency. However, musical signals have different distribution of frequency components than the synthetic data, therefore we had to adjust the parameters of the time-frequency representations. This led to a~change in the input dimension to $3~\times~480~\times~160$. These changes and the more challenging nature of the task led to slight modification of the architecture in comparison to the architecture in the experiment with synthetic data:

The number of kernels in the first three convolutional layers was raised to~64. The number of kernels in the last convolutional layer was raised to~16. The output dimension of this architecture was set to~6, as this was the number of classes. The batch size changed to~128 samples per batch. The number of weight updates was set to~11,000. To prevent unnecessary training, this set of experiments was set to terminate after~50 consecutive epochs without an improvement in validation loss as measured by categorical cross-entropy. The loss function and optimization algorithm remained the same as well as the used programming language, framework, and hardware. Experiments are fully reproducible and can be obtained by running the code in the repository~\cite{gs-gt}. Consider this repository also for more details about the networks architecture.

In this set of experiments, we have trained~10 models in total with five~scenarios with limited training set (5, 11, 55, 110, and 550~batches each containing~128 samples) for each time-frequency representation. In all of these scenarios, the sizes of the validation and testing sets remained at their full sizes each consisting of 188~batches containing 24,000~samples.

\subsubsection{Results}

\noindent
Table~\ref{table:goodsoundsresults} shows the accuracies of the model's best weight update on training, validation, and testing sets. The best weight update was chosen based on the performance on the validation set. As before, more details can be found in the aforementioned repository. In this experiment using GoodSounds data, a~similar trend as for the synthetic data is visible. GS performs better than GT if we are limited in training set size, i.e., having~640 training samples, the GS outperformed GT. 

\begin{figure}[H]
\includegraphics[width=\linewidth]{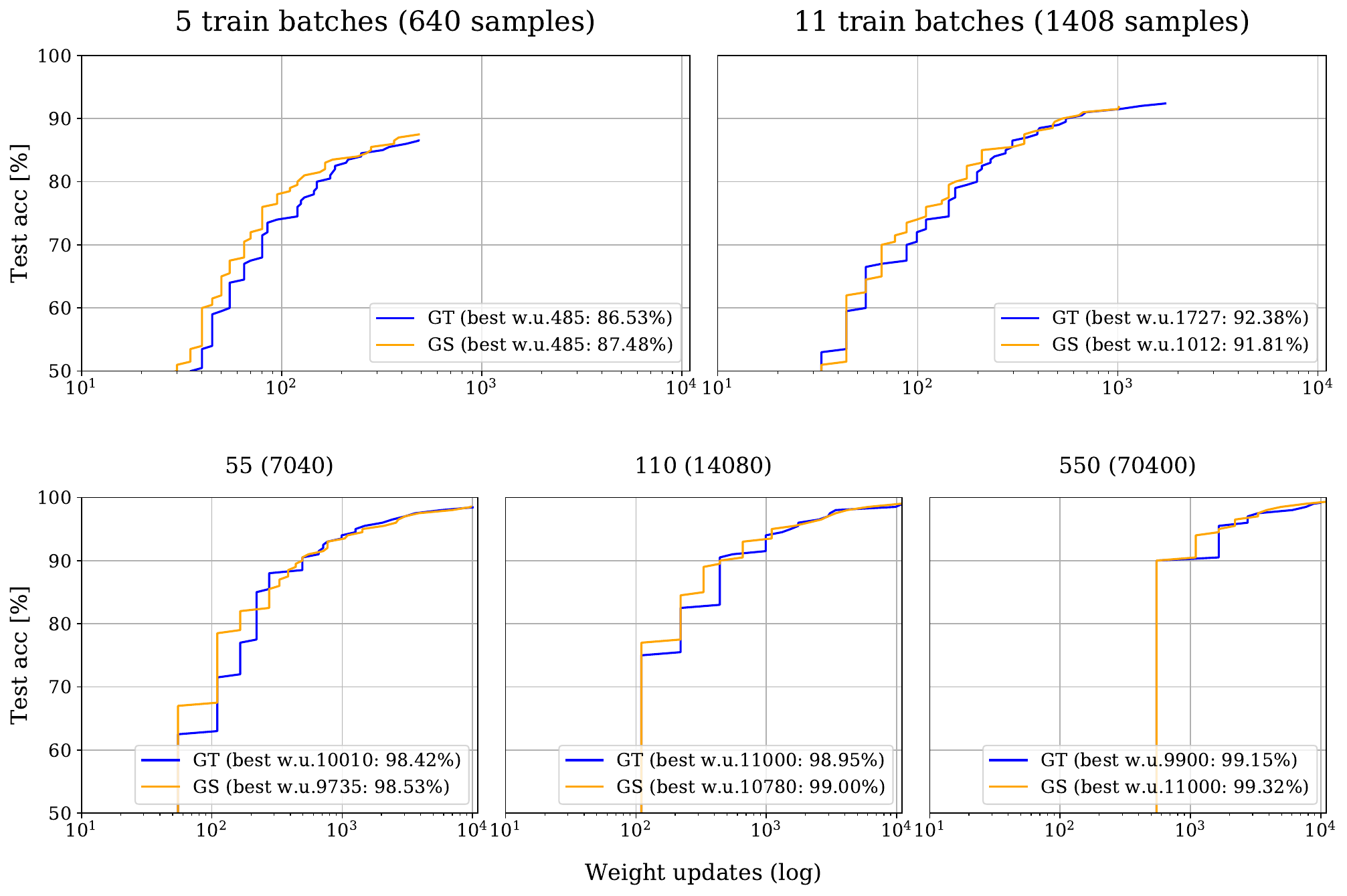}
\caption{CNN performance milestone reached over number of weight updates---GoodSounds data. Figure notation: 
Test acc---Accuracy performance metric measured on the testing set. 
Best w.u.---Weight update after which the highest performance was reached.}
\label{fig:performance-gs}
\vspace{5mm}
\end{figure}

\begin{table}[H] 
\centering
\caption{Performance of CNN: GoodSounds data.}
\scalebox{.85}[.85]{\begin{tabular}{cccccccc} 
\toprule 
\textbf{TF}	&	\textbf{N Train}	&	\textbf{N Valid}	&	\textbf{N Test}	&	\textbf{BWU}	&	\textbf{Train}	&	\textbf{Valid}	&	\textbf{Test	}\\
\midrule 
GS	&	640	&	24,000	&	24,000	&	485	&	0.9781	&	0.8685	&	0.8748	\\
GT	&	640	&	24,000	&	24,000	&	485	&	0.9766	&	0.8595	&	0.8653	\\
\midrule
GS	&	1408	&	24,000	&	24,000	&	1001	&	0.9773	&	0.9166	&	0.9177	\\
GT	&	1408	&	24,000	&	24,000	&	1727	&	0.9943	&	0.9194	&	0.9238	\\
\midrule
GS	&	7040	&	24,000	&	24,000	&	9735	&	0.9996	&	0.9846	&	0.9853	\\
GT	&	7040	&	24,000	&	24,000	&	8525	&	0.9999	&	0.9840	&	0.9829	\\
\midrule
GS	&	14,080	&	24,000	&	24,000	&	10,780	&	0.9985	&	0.9900	&	0.9900	\\
GT	&	14,080	&	24,000	&	24,000	&	9790	&	0.9981	&	0.9881	&	0.9883	\\
\midrule
GS	&	70,400	&	24,000	&	24,000	&	11,000	&	0.9963	&	0.9912	&	0.9932	\\
GT	&	70,400	&	24,000	&	24,000	&	8800	&	0.9934	&	0.9895	&	0.9908	\\
\bottomrule
\end{tabular}}
\begin{tabular}{@{}c@{}} 
\multicolumn{1}{p{\textwidth -.88in}}{\footnotesize Table notation: TF---Time-frequency representation. 
N train, N valid and N test---Number of samples in training, validation and testing sets. 
BWU---Weight update after which the highest performance was achieved on the validation set. 
Train, valid and test---accuracy on training, validation, and testing sets.}
\end{tabular}

\label{table:goodsoundsresults}
  
\end{table}
\noindent
In Figure \ref{fig:performance-gs}, we again see that in earlier phases of the training, GS reaches higher accuracies after less weight updates than GT. This effect diminishes with bigger training sets and vanishes in case of~550 training batches.

\section{Discussion and Future Work}
\noindent
In the current contribution, a~scattering transform based on Gabor frames has been introduced, and its properties were investigated by relying on a~simple signal model. Thereby, we have been able to mathematically express the invariances introduced by GS within the first two layers. 

The hypothesis raised in Section~\ref{DCNN}, that explicit encoding of invariances by using an adequate feature extractor is beneficial when a~restricted amount of data is available, was substantiated in the  experiments presented in the previous section. It was shown that in the case of a~limited dataset the application of  a~GS representation  improves the performance in classification tasks in comparison to using GT.

In the current implementation and with parameters described in Section~\ref{section:synthetic-data}, the GS is approximately 3~times more expensive to compute than GT. However, this transformation needs to be done only once---in the preprocessing phase. Therefore, the majority of computational effort is still spent during training, e.g., in the case of the GoodSounds experiment, the training with GS is ~2.5~times longer than with GT. Note that this is highly dependent on the used data handling pipeline, network architecture, software framework, and hardware, which all can be optimized to alleviate this limitation.
Although GS is more computationally-expensive, the obtained improvement justifies its use in certain scenarios; in particular, for classification tasks which can be expected to benefit from the invariances introduced by GS. In these cases, the numerical experiments have shown that by using GS instead of GT a~negative effect of a~limited dataset can be compensated. 

Hypothetically, with enough training samples, both GS and GT should perform equally assuming sufficient training, i.e., performing enough weight updates. This is shown in the results of both numerical experiments presented in this article (see Tables~\ref{table:synthresults} and \ref{table:goodsoundsresults}). This is justified by the fact that GS comprises exclusively the information contained within GT, only separated into three~different channels. We assume it is easier for the network to learn from such a~separated representation. The evidence to support this assumption is visible in the earlier phases of the training, where GS reaches higher accuracies after less weight updates than GT (see Figures \ref{fig:performance} and \ref{fig:performance-gs}). This effect increases with smaller datasets while with very limited data GS even surpasses GT in performance. 
This property can be utilized in restricted settings, e.g., in embedded systems with limited resources or in medical applications, where sufficient datasets are often too expensive or impossible to gather, whereas the highest possible performance is crucial.

We believe that GT would eventually reach the same performance as GS, even on the smallest feasible datasets, but the network would need more trainable parameters, i.e., more complex architecture to do the additional work of finding the features that GS already provides. Unfortunately, in such a~case, it remains problematic to battle the overfitting problem. This opens a~new question---whether the performance boost of GS would amplify on lowering the number of trainable parameters of the CNN. This is out of the scope of this article and will be addressed in the future work.

In another paper \cite{DBLP:journals/corr/abs-1903-08950}, we extended GS to mel-scattering (MS), where we used GS in combination with a~mel-filterbank. This MS representation reduces the dimensionality, and therefore it is computationally less expensive compared to GS.

It remains to be said that the parameters in computing GS coefficients have to be carefully chosen to exploit the beneficial properties of GS by systematically capturing data-intrinsic invariances.\\
Future work will consist of implementing GS on the GPU, to allow for fast parallel computation. At~the same time, more involved signal models, in particular, those concerning long-term correlations, will be studied analytically to the end of achieving results in the spirit of the theoretical results presented in this paper.

\section*{Acknowledgment}\noindent
 This work was supported by the Uni:docs Fellowship Programme for Doctoral Candidates in Vienna, by the Vienna Science and Technology Fund (WWTF) project SALSA (MA14-018), by the International Mobility of Researchers (CZ.02.2.69/0.0/0.0/16 027/0008371), and by the project LO1401. Infrastructure of the SIX Center was used for computation.


\bibliographystyle{unsrt}
\bibliography{references}

\end{document}